\documentclass[10pt,journal,twoside,compsoc]{IEEEtran}

\usepackage[utf8]{inputenc}
\usepackage[T1]{fontenc}

\ifCLASSOPTIONcompsoc
\else
	\usepackage{caption}
	\DeclareCaptionFont{ieeeblue}{\color{accessblue}}
	\DeclareCaptionLabelFormat{myformat}{\figcapfont{\textbf{#1} \textbf{#2}}}
\fi

\usepackage[caption=false,font=footnotesize]{subfig}
\usepackage{url}
\usepackage[dvipsnames]{xcolor}  
\usepackage{multirow}

\definecolor{gray}{gray}{0.9}

\usepackage{soul}
\definecolor{burgundy}{rgb}{0.5, 0.0, 0.13}
\definecolor{brown(web)}{rgb}{0.65, 0.16, 0.16}
\newcommand{\pending}[1]{\textcolor{BurntOrange}{#1}}
\renewcommand{\pending}[1]{#1}
\newcommand{\markup}[1]{\textcolor{blue}{#1}}
\renewcommand{\markup}[1]{}
\newcommand{\new}[1]{\textcolor{blue}{#1}}
\renewcommand{\new}[1]{#1}
\newcommand{\drop}[1]{\textcolor{red}{#1}}
\renewcommand{\drop}[1]{}

\usepackage[nocompress]{cite}

\usepackage{amsmath,amssymb,amsfonts}
\usepackage{algorithmic}
\usepackage{graphicx}
\usepackage{textcomp}

\begin{document}

\title{2.5D Root of Trust: Secure System-Level Integration of Untrusted Chiplets}
\author{Mohammed Nabeel, Mohammed Ashraf, Satwik
Patnaik,~\IEEEmembership{Graduate Student Member,~IEEE}, Vassos~Soteriou,~\IEEEmembership{Senior Member,~IEEE}, Ozgur Sinanoglu,~\IEEEmembership{Senior Member,~IEEE}, and\\ Johann Knechtel,~\IEEEmembership{Member,~IEEE}
\IEEEcompsocitemizethanks{
\IEEEcompsocthanksitem M.\ Nabeel, M.\ Ashraf, O.\ Sinanoglu and J.\ Knechtel are with the Division of Engineering, New York University Abu Dhabi, Abu Dhabi, 129188, UAE. Emails: \{mtn2, ma199, ozgursin, johann\}@nyu.edu 
\IEEEcompsocthanksitem S.\ Patnaik is with the Department of Electrical and Computer Engineering, Tandon School of Engineering, New York University (NYU), Brooklyn, NY, 11201, USA. Email: sp4012@nyu.edu
\IEEEcompsocthanksitem V.\ Soteriou is with the Department of Electrical Engineering, Computer Engineering and Informatics, Cyprus University of Technology, Limassol, Cyprus. E-mail:
vassos.soteriou@cut.ac.cy
\IEEEcompsocthanksitem Corresponding authors: Mohammed~Nabeel, Ozgur~Sinanoglu,
and Johann~Knechtel (e-mails: \{mtn2, ozgursin, johann\}@nyu.edu).}

\thanks{Copyright $\copyright$ 2020 IEEE. Personal use of this material is permitted.  However, permission to use this material for any other
	purposes must be obtained from the IEEE by sending an email to pubs-permissions@ieee.org. DOI: 10.1109/TC.2020.3020777}
}

\markboth{IEEE Transactions on Computers,~Vol.~X, No.~X, Month~202X}%
{Nabeel \MakeLowercase{\textit{et al.}}: 2.5D Root of Trust: Secure System-Level Integration of Untrusted Chiplets}

\IEEEtitleabstractindextext{%
\begin{abstract}
\textbf{Dedicated, after acceptance and publication, in memory of the late Vassos Soteriou.}
For the first time, we leverage the \textit{2.5D interposer technology} to establish system-level security in the face of hardware- and
software-centric adversaries.  More specifically, we integrate \textit{chiplets} (i.e., third-party hard intellectual property of complex
functionality, like microprocessors) 
using a security-enforcing interposer.
Such hardware organization provides a robust \textit{2.5D root of trust} for
trustworthy, yet powerful and flexible, computation systems.
The security paradigms for our scheme, employed firmly by design and construction, are:
1)~stringent physical separation of trusted from untrusted components, and
2)~runtime monitoring.
The system-level activities of all untrusted commodity chiplets are checked continuously against \textit{security policies}
via physically separated security features.
Aside from the security promises, the good economics of outsourced supply chains are still maintained; the system vendor is free to procure
chiplets from the open market,
while only producing 
the interposer and assembling the 2.5D system oneself.
We showcase our scheme using
the \textit{Cortex-M0} core and the \textit{AHB-Lite} bus by \textit{ARM},
building a secure 64-core system with shared memories.
We evaluate our scheme through hardware simulation,
considering different threat scenarios.
Finally, we devise a physical-design flow for 2.5D systems, based on commercial-grade design tools,
to demonstrate and evaluate our \textit{2.5D root of trust}.
\end{abstract}

\begin{IEEEkeywords}
Hardware Security,
2.5D Integration,
Active Interposer,
Chiplets,
Multi-Core System,
Runtime Monitoring,
Policies
\end{IEEEkeywords}
}

\maketitle

\IEEEraisesectionheading{
\section{Introduction}
\label{sec:introduction}
}

\IEEEPARstart{S}{ecurity} for computation systems
has focused traditionally on software aspects, and it is understood that related efforts have to remain ongoing.  Nowadays, the hardware itself
has also become susceptible to misuse.
Due to the economics of integrated circuit (IC) design and manufacturing, which dictates a production mode that is distributed across many
vendors, an adversary involved at any step
within the supply chain
may, e.g., pirate the design intellectual property
(IP)\new{~\cite{BRS17, knechtel19_IP_COINS}}.
Malicious modifications, also known as \textit{hardware Trojans},
could also be introduced during design, manufacturing or deployment, and can be stealthy and severe, e.g., see~\cite{
	yang16_a2}.
Besides, even imprudent decisions made by legitimate designers can
give rise to critical vulnerabilities, e.g., as
demonstrated by the
\textit{ZombieLoad}
attack~\cite{
	Schwarz2019ZombieLoad}.
These and other threats certainly impact the prospects for secure use of computation systems negatively.

\begin{figure}[tb]
\centering
\includegraphics[width=0.75\columnwidth]{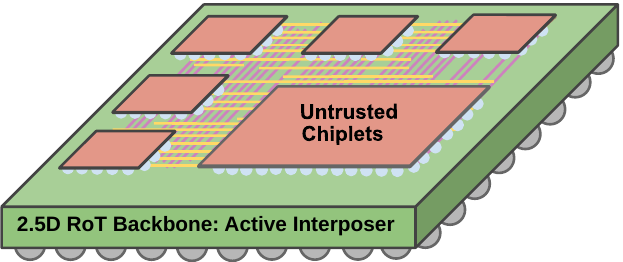}
\caption{An active interposer acts as backbone for the \textit{2.5D root of trust (RoT)}, which enables secure system-level integration of untrusted
	chiplets. All system-level communication is ``policed'' by the interposer.
\label{fig:concept}
}
\end{figure}

In this paper, we harness the opportunities offered by state-of-the-art 2.5D technologies for advancing hardware security.  That is, we extend
the scope for modern computation systems by means of a robust, system-wide, and hardware-enforced security scheme
that is enabled by 2.5D design and construction (Fig.~\ref{fig:concept}).
Next, we discuss the background and motivation for our work in more detail.

\subsection{Hardware Security Features}
\label{sec:HWSFs}

There exist many hardware security features (HWSFs), seeking to
mitigate various software- and/or hardware-based threats at runtime. They include enclaves
for trusted execution, like the industrial \textit{ARM TrustZone} and \textit{Intel SGX} or the academic \textit{MIT Sanctum} (these and others are reviewed in~\cite{maene18}),
wrappers for monitoring and cross-checking of untrusted third-party intellectual property (IP)
modules~\cite{basak17%
}, centralized IP infrastructures for secure system design~\cite{wang15_IIPS},
verification of computation~\cite{wahby16},
secure task scheduling~\cite{liu15_TETC},
secure network-on-chip (NoC) architectures~\cite{fiorin08%
}, \textit{etc.}
Besides, there are also design-time mitigation schemes, e.g., using high-level synthesis strategies for detection, collusion prevention,
and isolation of malicious IP~\cite{rajendran16}.

\begin{figure*}[tb]
\centering
\includegraphics[width=.99\textwidth]{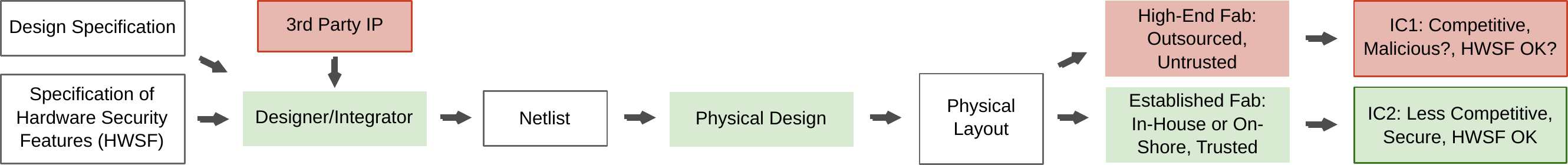}
\caption{%
IC supply chain with focus on hardware security features (HWSFs).
Green and red boxes represent trusted and untrusted entities and assets, respectively.
Implementing a trustworthy \textit{and} competitive IC
requires a trusted \textit{and} high-end fabrication process, two aspects that conflict with each other\new{, as also indicated in~\cite{BRS17,mavroudis17}}.
Few, if any, of the outsourced high-end facilities may be considered trustworthy, whereas maintaining an advanced, trusted facility
on-shore or in-house is too costly in practice.
\label{fig:security_motivation}%
}
\end{figure*}

Notwithstanding their good prospects, most, if not all, HWSFs eventually suffer from tailored attacks, e.g., see~\cite{sec17-lee-jaehyuk} which exploits a memory
corruption vulnerability in the enclave software of \textit{Intel SGX}.
In general, HWSFs arguably form prime targets---strategic adversaries would first aim to bypass or disable them so that
further attacks
can thereafter remain ``under the radar.''
Thus, HWSFs become especially vulnerable once adversaries can tamper with the outsourced IC supply chain
(Fig.~\ref{fig:security_motivation}).\new{ (For other security concerns related to outsourced IC supply chains, see also,
	e.g.,~\cite{knechtel19_IP_COINS,BRS17,rajendran16}.)}
We note that the above circumstance also imposes an important practical challenge, namely how to implement ICs that are
high-end, competitive, and relatively cheap,
\textit{yet} trustworthy and secure.
This is because trusted manufacturing facilities (be they either in-house or on-shore and certified) typically cannot offer 
the latest high-end technology nodes, as doing so would be too costly.

To the best of our knowledge, \textit{none of the proposed HWSFs in prior art
can fully withstand malicious modifications}.
Once HWSFs are permanently tampered with, even when such malicious activity is subsequently detected,
the resulting loss of security guarantees cannot be restored---the
chips become untrustworthy and as such possibly even unusable.
We note that there are efforts to render hardware secure in the direct presence of Trojans;
such schemes typically leverage some formalism like multi-party computation~\cite{bronchain18
} or verification and
proofing~\cite{wahby16}.
While promising, such schemes still require that at least some parts
of the system remain trustworthy, i.e., that some parts are guaranteed to be free of any malicious modifications.
Moreover, such schemes are less applicable to general-purpose, high-performance computation systems, as
the underlying formalism requires extensive system- and circuit-level support, which naturally also tends to impose considerable overheads.

\subsection{2.5D and 3D Integration}
\label{sec:2.5D_3D}

The umbrella of 2.5D and 3D integration technologies collectively embrace the notion of ``building skyscrapers and city clusters of
electronics''~\cite{knechtel17_TSLDM}.
There exist two main drivers for 2.5D and 3D integration:
1)~the CMOS scalability bottleneck, which becomes more exacerbated for advanced nodes by issues like routability, pitch scaling, and process
variations;
2)~the need to advance heterogeneous and system-level integration.
Both drivers are also known as ``More Moore'' and ``More than Moore,'' respectively.
Next, we provide a brief overview on the related technologies.

Native 3D integration means to vertically stack and interconnect multiple chips or active layers.
This approach can be further classified by the underlying technology, with the main ones being:
1)~through-silicon via (TSV)-based 3D ICs,
2)~face-to-face (F2F) 3D ICs,
3)~monolithic 3D (M3D) ICs~\cite{knechtel17_TSLDM}.
Various studies, prototypes, and commercial products have shown that native 3D integration can indeed offer significant benefits over conventional 2D
ICs, e.g., see~\cite{aly19,gomes20
}.

   2.5D integration, also known as the \textit{interposer technology}, facilitates system-level integration of
2D chips side-by-side. An interposer
serves as an integration carrier and accommodates an underlying system-level interconnect
fabric to provide inter-chip communication~\cite{lau11,
	stow17,
	vivet20,
	kim19},
thereby resembling a modern version of a printed circuit board.
Building an advanced electronic system using an interposer is considered less complex than native 3D
integration~\cite{stow17,
	lau11}.
In fact, interposer-based systems are already established in the market, e.g., with the \textit{AMD Fiji} GPU system~\cite{lee16}
or \textit{Xilinx}'s \textit{Virtex-7} FPGAs~\cite{dorsey10}.

Besides classical passive interposer, comprising only passive components and wiring, a more promising option are \textit{active interposer} which
additionally incorporate logic.
Active interposers have been successfully demonstrated, e.g., see~\cite{takaya13,
vivet20,hellings15}.
Active interposers are preferably implemented using mature technology nodes, i.e., for commercial cost savings, yield, and, even more importantly for our
work, ease of access to
an established and trusted facility.
Regarding manufacturing and integration cost, we note that both sides have been argued for, i.e., interposers are 
cheaper than 3D ICs~\cite{lau11} versus interposers are more costly~\cite{velenis13}.
However, once system-level cost are considered, the interposer technology remains promising, also because active interposers improve
testability~\cite{takaya13,hellings15,vivet20} and, thereby, allow to better manage yield of the final system.

Remarkably, the trend toward 2.5D and 3D integration can serve well to advance various notions of hardware security.
Still, one has to account carefully for related limitations.
For example, Valamehr {\em et al.}~\cite{
	valamehr13} propose a runtime monitor to be 3D-stacked
on top of a commodity processor, along with dedicated HWSFs that allow for tapping and rerouting of sensitive signals.
However, all these HWSFs rely on introspective interfaces within the commodity processor.
Therefore, when subjected to an untrusted supply chain, that scheme could fail entirely
once these interfaces in the commodity processor are tampered with.
The authors themselves acknowledge this significant limitation in~\cite{valamehr13}.

We provide a high-level overview on ours and selected prior works in Table~\ref{tab:prior_art}, and
we discuss the motivation and contributions of our work in more detail in Sec.~\ref{sec:motivation}.

\begin{table}[tb]
\centering
\scriptsize
\caption{Selected Works Leveraging 2.5D/3D Integration for Hardware Security}
\label{tab:prior_art}
\setlength{\tabcolsep}{1.8mm}
\begin{tabular}{cccccc}
\hline
\textbf{Reference} & 
\textbf{Style} & 
\textbf{Security Scope; Means} & 
\textbf{Trusted Asset} \\ 
\hline 
\multirow{2}{*}{\cite{valamehr13}} & \multirow{2}{*}{TSV} & Runtime monitoring; & \multirow{2}{*}{Whole 3D IC} \\
& & split manufacturing (SM) & \\ \hline
\cite{xie17} & 2.5D & IP protection; SM & Passive interposer \\ \hline
\cite{imeson13} & 2.5D & Trojan prevention; SM & Passive interposer \\ \hline
\cite{yan17_camo} & M3D & IP protection; camouflaging & Whole 3D IC \\ \hline
\multirow{2}{*}{\cite{patnaik19_3D_TETC}} & \multirow{2}{*}{F2F} & IP protection, Trojan prevention; & \multirow{2}{*}{Only BEOL} \\
& & SM, camouflaging & \\ \hline
\cite{knechtel17_TSC_DAC}, & \multirow{2}{*}{TSV} & Side-channel mitigation; & \multirow{2}{*}{Whole 3D IC} \\
\cite{bao17} & & 3D integration by itself & \\ \hline
\multirow{3}{*}{\textbf{Ours}} & \multirow{3}{*}{\textbf{2.5D}} & \textbf{Security by construction,} & \multirow{3}{*}{\textbf{Active interposer}} \\
& & \textbf{runtime monitoring;} & \\
& & \textbf{stringent physical separation} & \\ \hline
\end{tabular}
\end{table}

\subsection{Chiplets: System-Level IP Integration}
\label{sec:chiplets}

Concurrently, there exist efforts for driving the notion of IP reuse toward the system level.
Under these efforts,
not only IP modules are to be reused at the chip level, but rather entire \textit{chiplets} at the system level.
Chiplets are relatively small chips encapsulating certain levels of complex functionality, like 
a microprocessor, as hard physical IP.
The potential benefits of using chiplets are
lower design and manufacturing costs, improved yield through separating technologies, and greater design flexibility.
As such, the economic benefits, especially for small-volume development of heterogeneous and large-scale systems, are becoming enormously promising.
It is only logical that the 2.5D interposer technology is at the heart of these efforts.

A prominent initiative for chiplets integration is \textit{DARPA's Common Heterogeneous Integration and Intellectual Property Reuse Strategies
(CHIPS)}~\cite{CHIPS}.
The main objectives for CHIPS are: 1)~realize a modular design process and manufacturing flow, and
2)~establish standards for physical interfaces.
Ultimately, the goal is to achieve ``plug-and-play integration'' of large-scale and heterogeneous systems, as
opposed to the traditional, monolithic flow for 2D ICs.
In general, chiplets integration has been well-received by both the academia (e.g., see~\cite{yin18,stow17,
	kim19}) and the
industry, with relevant products and technologies already in the market, e.g., see the \textit{AMD Fiji}
system~\cite{lee16}
and \textit{Intel's Embedded Multi-Die Interconnect Bridge (EMIB)}~\cite{
	EMIB19}.

\subsection{Motivation and Contributions}
\label{sec:motivation}

In this paper, we harness the opportunities offered by state-of-the-art 2.5D technologies for advancing hardware security.
More specifically,
	we propose the assembly of:
1)~potentially untrusted commodity chiplets and memories, and 
2)~physically separated, entrusted communication interfaces and HWSFs residing in an active interposer.
We refer to the resulting system in general and the security-enforcing interposer in particular as \textit{2.5D root of trust (RoT)} (Fig.~\ref{fig:concept}).

From a commercial point of view, we note that the system vendor has to design, produce, and
sell such 2.5D RoT systems. Here, the good economics of chiplets reuse are still maintained; that vendor has to manufacture only
a fraction of the overall system, namely the security-enforcing interposer (with the help of some established and trusted fabrication facilities)
and would then integrate the high-end but untrusted chiplets on top of that interposer (with the help of in-house or certified on-shore
packaging facilities).
Therefore, the final system establishes security and can offer good performance at a reasonable
cost---the crux illustrated in Fig.~\ref{fig:security_motivation} can thus be resolved.

At this point, one might wonder about using a passive interposer for an alternative, potentially less costly implementation of an 2.5D RoT, but
we argue that doing so would entail two key limitations.
First, scalability would be compromised. This is because all HWSFs would have to be implemented within one or multiple, dedicated
security chiplet(s), which would then become the ``bottleneck'' for system-level communication through the interposer.
In fact, optimizing interposer interconnects is an area of research by itself,
	where active interposers are considered promising as well~\cite{akgun16,
yin18}.
Second, the need for trustworthy manufacturing of an interposer \textit{and} some security chiplet(s)
might well undermine the good economics of the scheme.
In short, we advocate for an active interposer for our proposed 2.5D RoT.

This paper makes the following contributions:
\begin{itemize}
\item We propose a novel \textit{2.5D root of trust} concept that, for the first time, establishes stringent physical
separation at the system level, between 1)~commodity chiplets and 2)~HWSFs residing in an active interposer.
In addition to \textit{ruling out common threat scenarios directly by construction}, the purpose of this concept is to \textit{enable continuous
	runtime monitoring of the system-level communication of all commodity chiplets}.
\item  Following the 2.5D RoT concept, we showcase a secure multi-core architecture with a system-level interconnect fabric and shared
memories.
We implement our scheme using the \textit{Cortex-M0} core and the
\textit{AHB-Lite} bus system, both by \textit{ARM}.
We develop dedicated HWSFs for memory access and data control that form an integral part of our scheme.
We release the license-free parts of our proof-of-concept (PoC) 64-core implementation to the community~\cite{webinterface}.
\item We develop an end-to-end physical-design flow for our 2.5D RoT, based on commercial tools. Our flow serves to design the active interposer
and supports a flexible design mode for chiplets procured as soft or hard IP.
Using this novel flow, we elaborate on the layout costs of our scheme in detail.
\item We evaluate our scheme
against various relevant attack scenarios.
We implement related security-enforcing policies and demonstrate them in action
against malicious runtime behavior, using a commercial hardware simulation workflow.
\end{itemize}

\section{Threat Model and Concept}\label{sec:concept}

\textit{Our concept does not require any trust assurance concerning the design and manufacturing of commodity chiplets}. In fact, we even assume
\textit{a priori} that chiplets do run malicious code and/or incorporate Trojans.

Crucially, such threats cannot undermine or compromise the system-level security of our scheme.
This is due to the fact that our 2.5D RoT scheme imposes physically and inevitably that any untrusted component has to depend on
the security-enforcing interposer for system-level communication,
whereas the trustworthiness and robust operation of that interposer 
are not subject to those untrusted components.
We note that this is in contrast to
most prior art where HWSFs are embedded monolithically in the same chip and, thus, remain subject
to the trustworthiness---or rather lack thereof---of all the related design and manufacturing stages.

\begin{figure}[tb]
\centering
\includegraphics[width=.95\columnwidth]{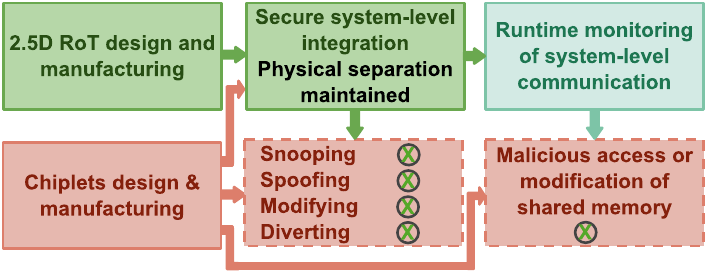}
\caption{2.5D RoT design and manufacturing stages (green), threat sources (red),
	and runtime monitoring (turquoise).
		It is imperative that the active interposer is designed and manufactured by trusted parties.
\label{fig:SecurityFlow}
}
\end{figure}

Our threat model and concept are illustrated in Fig.~\ref{fig:SecurityFlow}, with details for both discussed next.

\subsection{Security Threats and Assumptions}
\label{sec:SecurityThreats}

When seeking to securely integrate various components at the system level, different threats are to be considered,
which concern the system-level communication and all involved components~\cite{basak17}.
More specifically, in our work, a malicious chiplet may exercise the following attacks: 
\begin{enumerate}
\item Passive reading, also known as \textit{snooping}, i.e., a chiplet illicitly reads or gathers data that is meant for/authorized to
	other chiplets;
\item Masquerading, also known as \textit{spoofing}, i.e., a chiplet disguises or poses itself as another one, to
	illicitly control services or request data from other chiplets;
\item Modifying, i.e., a chiplet maliciously changes the data exchanged legally between other
	chiplets;
\item Diverting, i.e., a chiplet maliciously diverts the data exchanged legally between two chiplets to a third, unauthorized chiplet; and/or
\item Man-in-the-middle, i.e., a chiplet ``hijacks'' the communication between two chiplets---this attack is closely related to all four above.
\end{enumerate}
As we focus on a multi-core architecture with shared memories,
we also have to consider another threat: 
\begin{enumerate}\addtocounter{enumi}{5}
\item Malicious accesses and modifications of shared-memory-resident data.
\end{enumerate}

We assume that any of these six threats can be introduced by: a)~untrusted components/chiplets---either unintentionally via ``design bugs'' or
intentionally via Trojans---or b)~malicious software running on the cores.

We assume that any attack is exercised through system-level communication across chiplets.
Therefore, any adversarial activities conducted \textit{within} chiplets, such as covert channels across cores (e.g., \cite{masti15}),
side-channel- or fault-driven attacks across cores and their caches/buffers (e.g., \cite{osvik05,
Schwarz2019ZombieLoad}), or fault injection on privileged hardware interfaces (e.g., \cite{murdock20})
are all considered out of scope for this work.
Furthermore, we assume a trusted runtime environment.
Thus, any threats like side-channel or physical fault-injection attacks conducted by malicious end-users (e.g., \cite{brier04%
})
are considered out of scope as well.

Finally, we assume that the design and fabrication of commodity chiplets is outsourced and, hence, untrusted, whereas
the design and manufacturing of the 2.5D RoT and the system-level assembly
    must all be carried out in a trusted environment.
This also means that, in this work, we do \textit{not} seek to detect or prevent Trojans within chiplets (recall that we rather assume \textit{a priori} that
Trojans are present in the untrusted chiplets). Given that related techniques (e.g.,
\cite{lyu2019efficient,chakraborty2009mero}) are orthogonal to our efforts, such techniques could still be 
leveraged,
	to render the final system even more robust to begin with.

We note that, among other scenarios, \textit{DARPA's CHIPS}
	and \textit{DARPA's Secure Processing Architecture by Design
	(SPADE)}
programs both match well with said assumptions.
This is because government agencies seeking to build small numbers of large-scale, heterogeneous systems
in a cost-efficient manner are advised to
utilize chiplets which, when obtained from the open market, are potentially malicious.
To ensure secure computation nevertheless, within a trusted runtime environment readily enforceable by the agencies,
schemes like ours become essential.

\subsection{Security Concept and Working Principles}\label{subsec:SecurityConceptPrinciples}

\begin{figure}[tb]
\centering
\includegraphics[width=0.8\columnwidth]{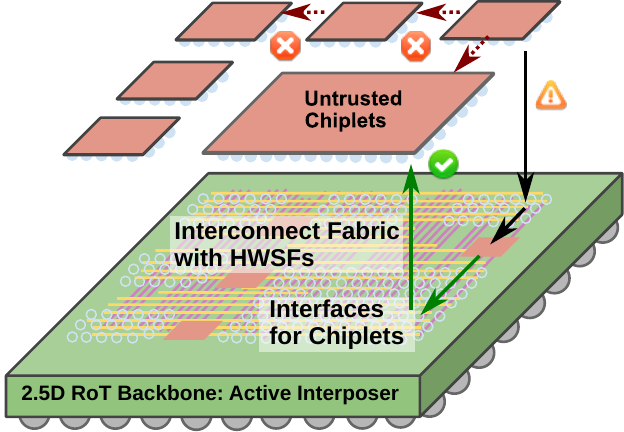}
\caption{Our scheme prevents common threats like snooping by construction. System-level communication must utilize the 2.5D RoT backbone
	and, thus, cannot be tampered with by other chiplets (red crosses). This is because the interposer's interfaces, which
	each chiplet is physically attached to, enforce that any communication request (yellow warning sign) is passed to and controlled by the HWSFs
	residing in the active interposer.
	Only approved communication (green tick) is passed on.
\label{fig:concept_extended}
}
\end{figure}

Our scheme is the first, to our best knowledge, that \textit{rules out the above threats 1)--5)
by design and construction.}
As Fig.~\ref{fig:concept_extended} illustrates, this kind of built-in security occurs as the system-level interconnect fabric with all its
interfaces and HWSFs are physically separated from the untrusted components. Therefore,
components/chiplets remain completely unaware of and isolated from any communication not directly addressed to or created by them.
For example, regarding \textit{spoofing}, we realize a hard-coded assignment of component identifiers (IDs) directly via the interconnect interfaces
which are residing exclusively in the 2.5D RoT. Thus, a malicious component cannot masquerade itself as another in the first place.

\begin{figure*}[tb]
\centering
\includegraphics[width=.65\textwidth]
{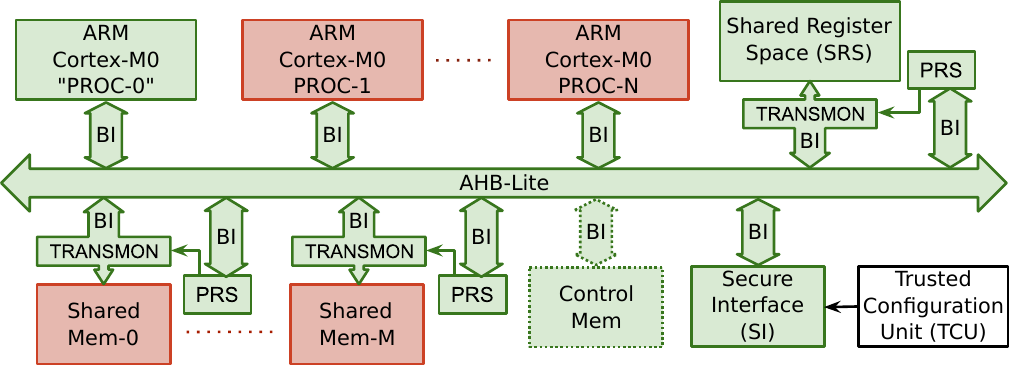}
\caption{\markup{Figure revised---}Block diagram for the proposed ISEA architecture, where components shown in green are security-critical and implemented exclusively in the
2.5D RoT.
Components shown in red constitute untrusted commodity components, implemented in various chiplets.
Note that all the bus interfaces (BIs) are also implemented in the 2.5D RoT. Among other considerations, this serves to ensure that master IDs for any
transaction emanating from some core cannot be tampered with by any of the other cores.
Note that ``PROC-0'' is a trustworthy core which is separate from all the cores in the commodity chiplets; it is implemented entirely in the 2.5D RoT and reserved for
scheduling and other management tasks such as compilation and updating of security policies residing in the PRSs (policy register spaces).
For the optional memory-security features (residing in TRANSMONs, not illustrated separately here), the related control memory is to be provided as a
separate, trusted memory chiplet.
\label{fig:architecture}
}
\end{figure*}

Furthermore, we utilize the notion of \textit{security policies}
for runtime monitoring against malicious access or modification of the system-level shared memory.
To do so,
we devise HWSFs that allows us to enforce a fully-controlled memory access scheme.
We apply stringent principles as follows, with
related technical details provided in Sec.~\ref{sec:ArcImplementation}.

First, any memory access not explicitly allowed for, via some policy, is denied by default.
Second, the continuous ``policing'' of memory access incurs uniform latency,
independent of whether access is allowed or denied, and
any denied access is responded to with a generic \textit{error message}.
These principles in conjunction
ensure that an adversary cannot infer whether the requested region is protected
or not used at all, which may serve well to hinder any related side-channel inference.
Third, to protect against faults and malicious data modifications within memories themselves, we
advocate for optional memory-security features like error correction codes (ECCs).
Here, the actual data and the results of the security features
are to be stored in physically separate locations and
cross-checked upon reading.

In short, common threats are ruled out by construction, malicious memory accesses are blocked, and erroneous data is rejected; all these
security principles are enabled directly at the 2.5D RoT.
For any such adversarial case, the overall system may experience a (temporary) loss of functionality or data, but its \textit{integrity and trustworthiness remain intact},
   an outcome which constitutes the main focus of this work.

\section{Architecture of 2.5D Root of Trust}\label{sec:ArcImplementation}

Here we describe the architecture of our 2.5D RoT, called \textit{\underline{I}nterposer-based~\underline{S}ecurity-\underline{E}nforcing
\underline{A}rchitecture}, or \textit{ISEA} for short.
The key paradigms of ISEA are: 1)~to physically separate commodity components (chiplets in our case) from the HWSFs and 2)~to monitor any memory-related,
system-level communication
at runtime.
In more colloquial terms, one can memorize the term ISEA as ``I see ya,'' which reflects upon the
idea of continuous system-level monitoring.

The novelty and enabler for ISEA is the security-enforcing active interposer,
which serves as integration carrier and as ``physical barrier'' for any communication-centric security fallacies to propagate through the system.
Toward that end, the interposer hosts the
system-level interconnect fabric along with all proposed HWSFs.
Therefore, any communication emanating from untrusted chiplets is inevitably handled and controlled by the interposer.
More specifically, in this work, we focus on \textit{shared-memory transactions} initiated by cores residing within chiplets.
The legality of any such transaction is verified using
various kinds of security policies; details and examples for such policies are provided further below and in Sec.~\ref{sec:SecurityAnalysis}.

\subsection{System Implementation}

\subsubsection{Overview}\label{sec:ImplOverview}

Figure~\ref{fig:architecture} depicts the block diagram for the proposed ISEA architecture.
Key to ISEA are \textit{\underline{Trans}action~\underline{Mon}itors (TRANSMONs)} which administer the various policies; the functionalities and implementation of TRANSMONs and
all other HWSFs are explained in detail further below.

In this paper, without loss of generality,
we consider
the \textit{ARM Cortex-M0} core for the commodity chiplets.
For the system-level interconnect fabric,
we leverage the \textit{ARM \underline{A}dvanced \underline{H}igh Performance \underline{B}us \underline{Lite} (AHB-Lite)}.
AHB-Lite facilitates communication among bus-attached master components (cores in our case) that initiate transactions and bus-attached slave
components (system-level shared memories in our case) that respond to these requests.
AHB-Lite transfers data values, addresses, and control info;
it is managed by components such as arbiters, decoders, multiplexers, \textit{etc.},
all of which collectively implement the AMBA protocol
(\textit{\underline{A}dvanced \underline{M}icrocontroller \underline{B}us \underline{A}rchitecture}).
We choose AHB-Lite as it is technology-independent, widely used
in the industry,
and encourages modular design,
all while offering high performance.
Note that AMBA provides a secondary bus which functions as a slave to AHB-Lite,
called the \underline{A}dvanced
\underline{P}eripheral \underline{B}us (APB), used for lower-bandwidth peripheral devices such as I/O ports. While APB could also be incorporated into 
ISEA, here we focus on AHB-Lite.

We release the license-free parts of our PoC implementation to the community~\cite{webinterface}.
We exhibit only a particular instance of ISEA here;
our scheme can be easily retro-fitted to secure other
systems,
with different chiplets, cores, and/or interconnects.
This is because the key principles of our scheme are agnostic to these implementation aspects.

We note that Cortex-M0 does not provide a cross-communication interface; hence, direct M0-to-M0 communication is not possible, thereby also
excluding such direct message-passing for this PoC implementation of ISEA.
Still, ISEA could be extended toward direct message-passing; TRANSMONs
would be incorporated directly in-between the components to be monitored. This approach would
also be applicable for, e.g., traditional system-on-chip (SoC) designs where IP cores may be connected via direct links.
Besides, modern architectures may also contain hardware accelerators which can use integrated memories and/or
external, shared memories. For the latter, the accelerators have to act as a bus master, like all other cores, to access those shared memories.
Thus, ISEA can also be used to monitor transactions by other components, like accelerators, not just regular cores.
Finally, ISEA could also be extended toward other types of system-level fabrics, like NoCs.

We emphasize again that it is essential for ISEA that the system-level interconnect fabric,
its interfaces,
and all proposed HWSFs are implemented exclusively in the active
interposer, thereby constituting the 2.5D RoT by design and construction.
For example, all communication requests passed onto the AHB-Lite bus system are associated with a master ID, whose
assignment is handled by the bus-interface ports the chiplets are physically attached to---for ISEA, these ports are implemented in the trusted interposer, not within the chiplet.
Thus, concerning spoofing, by construction there is no attack surface that could be leveraged by some Trojan or malicious
software running within the chiplets and seeking to alter the master IDs.
This and other scenarios are also illustrated in Fig.~\ref{fig:ISEA_examples}, Sec.~\ref{sec:SecurityAnalysis}.

\subsubsection{Our Hardware Security Features}

ISEA constitutes the following HWSFs:
\begin{enumerate}
\item TRANSMONs, along with their \underline{P}olicy \underline{R}egister \underline{S}paces (PRSs) to store the various policies;
\item a \underline{S}hared \underline{R}egister \underline{S}pace (SRS);
\item an ARM Cortex-M0 core called ``PROC-0''; and
\item the \underline{S}ecure \underline{I}nterface (SI).
\end{enumerate}
The purpose of these features and their components is explained next.

1)~A TRANSMON controls all transactions related to its attached memory chiplet, based on the policies stored in its PRS.
A TRANSMON itself comprises three or four components:
a)~the \textit{\underline{A}ddress \underline{P}rotection \underline{U}nit (APU)},
b)~the \textit{\underline{D}ata \underline{P}rotection \underline{U}nit (DPU)},
c)~the \textit{\underline{S}lave \underline{A}ccess \underline{F}ilter (SAF)},
and, optionally, d)~a memory-security feature.

All components establish security collectively, with their functionality elaborated in Sec.~\ref{sec:TRANSMON}. In a nutshell, the APU protects against undefined and/or unpermitted memory accesses,
the DPU protects against illegal data modification or leakage of restricted data into
the system's shared memory space,
the SAF serves to forward or reject requests which are approved or rejected, respectively,
and the memory-security feature serves to detect faults and malicious modifications within the memories themselves.

2)~The SRS can be used for secure data sharing, e.g., for \textit{semaphores}.  Although the SRS is implemented in the interposer, a TRANSMON and
its related PRS are still required, to realize access control and runtime monitoring.

3)~The interposer-embedded (and thus fully trustworthy) PROC-0
serves for scheduling and controlling the distributed computation, with commodity cores in the untrusted chiplets being
allocated and interrupted by PROC-0 at runtime as needed.
PROC-0 will further serve for mapping the system-level shared memory spaces, and for compiling and updating the application-specific sets of
policies residing in the PRSs.
It is important to note that PROC-0 does not constitute a ``bottleneck''
as it is \textit{not} involved in each and every AHB-Lite transaction, but it is only used in exercising this kind of system-level management.

4)~An external \underline{T}rusted \underline{C}onfiguration \underline{U}nit (TCU)
is responsible
for loading the application(s) and initial data onto the system,
and for retrieving the final results from the system.
All these tasks are performed using the SI, which has privileged access to the AHB-Lite. Recall that we assume a
trusted runtime environment; attacks misusing the TCU or SI in the field are thus out of scope.
In any case, access to the TCU or SI can by protected by cryptographic primitives.

\subsection{Transaction Monitor (TRANSMON)}
\label{sec:TRANSMON}

\begin{figure}[tb]
\centering
\includegraphics[width=\columnwidth]
{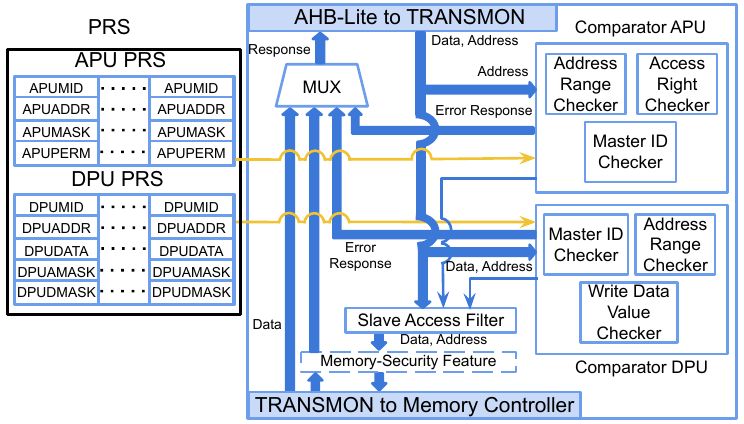}
\caption{Block diagram for the TRANSMON's micro-architecture.
See also Sec.~\ref{sec:TRANSMON} for technical details as well as Figs.~\ref{fig:APU_example}, \ref{fig:DPU_example} for examples of APU, DPU
policies in action, respectively.
\label{fig:TRANSMON}
}
\end{figure}

Key to ISEA's operation are TRANSMONs, with their micro-architecture illustrated in Fig.~\ref{fig:TRANSMON}.
Note that we choose to place an individual TRANSMON in-between every memory slave and the AHB-Lite bus interface (Fig.~\ref{fig:architecture}).
While another option would be to place TRANSMONs in-between all the core masters and their respective bus interfaces, our design
decision offers two important benefits. \pending{First, a TRANSMON connected to a master would require additional
address bits decoding and checking (for the base address), which is already covered by AHB-Lite itself, whereas a TRANSMON connected to a
slave only requires decoding and checking for the offset address.} Second, a TRANSMON connected to a slave
enables us to keep track of the security policies relevant to only that slave, thereby helping with efficiency.

\subsubsection{TRANSMON Design: Overview, Working Principles}\label{subsec:TRANSMONDesign}

As described, a TRANSMON comprises an APU, a DPU, an SAF, optionally a memory-security feature, and some glue logic.
The APU and DPU each have access to their own PRS.
For efficiency, every
PRS is implemented using flip-flops.
Each APU PRS entry defines one APU policy concerning
some particular region in the system's shared memory space, physically allocated in the memory slave connected to that TRANSMON;
each DPU PRS entry defines one DPU policy concerning
some particular data.
Both APU and DPU policies are discussed in more detail below and examples are illustrated in Fig.~\ref{fig:APU_example} and Fig.~\ref{fig:DPU_example}.

TRANSMONs block all read or write requests that are violating any of their APU/DPU policies.
By default, TRANSMONs also block requests that cannot be matched to any policy, protecting the system against all such ``stray requests.''
Policy verification also involves the checking of master/slave IDs. 
In this context, as we implement the interconnect fabric and all its
interfaces physically exclusively in the active interposer, recall that \textit{there cannot, a priori, be any spoofing of
IDs, snooping, modifying, or diverting of data, or man-in-the-middle attacks}.
In case a request is blocked, the related TRANSMON passes an error message to the master which initiated the transaction
and an interrupt to the trusted PROC-0.
The memory access itself is then dropped by the SAF---it is thus guaranteed to never reach the memory.

As indicated, the trusted PROC-0 within ISEA serves for mapping the system-level shared memory spaces, and for compiling and updating
the application-specific sets of policies for each TRANSMON (more specifically, for its PRS).  For different applications running on the
system, depending on the scheduling,
policies can also be devised for protection of independent data sets of multiple applications running in parallel. Once a particular application
run is finalized, before dropping the related policies, PROC-0 should also clear the related memory regions, to avoid any posterior leakage of sensitive data.
Moreover, the trusted end-user is free to implement software-level analysis and management of all blocked requests.
Such management schemes may also decide whether masters which repetitively trigger requests to be blocked should be isolated completely from the system (by updating
the policies accordingly), in order to mitigate potential denial-of-service attacks.
In any case, software implementation for such ``bootstrapping'' and system-level management procedures
are scope for future work; in this work, we focus on the ISEA architecture, the implementation and physical design of all its HWSFs,
and on a security analysis based on functional hardware simulation runs.

\subsubsection{TRANSMON Design: Address Protection Unit (APU)}

The APU forms an integral part of the TRANSMON;
it serves to check all read or write memory requests.
As such,
full access control over all shared-memory ranges is exercised.
We design and implement the APU such that policy checking is 
acting during the address phase of the AHB-Lite protocol,
thereby avoiding additional cycle delays.

\begin{figure}[tb]
\centering
\includegraphics[width=\columnwidth]
{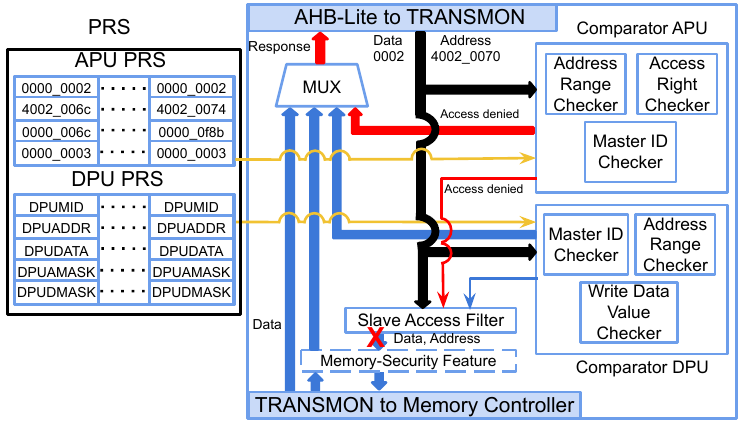}
\caption{APU policies in action in the TRANSMON.
Essentially, the core with ID 0x2 tries to access a memory region outside of the allowed ranges defined in the two policies. Therefore, the
	access is blocked by the SAF, Slave Access Filter, from passing to the memory controller, and an error message is returned.
See also Sec.~\ref{sec:SecurityAnalysis} for more details.
\label{fig:APU_example}
}
\end{figure}

Recall that each APU makes use of its own PRS to hold the polices related to its physically assigned memory slave.
As Fig.~\ref{fig:TRANSMON}
shows, an APU policy comprises four parameters:
\begin{itemize}
\item APUMID, which identifies the master allowed to initiate the particular memory request described by this policy;
\item APUADDR, a 32-bit memory address;
\item APUMASK, a 32-bit address mask; and
\item APUPERM, the access permission, i.e., whether read-only, write-only, or read-write.
\end{itemize}
An example for an APU policy is illustrated in Fig.~\ref{fig:APU_example}, and the related simulation is provided in Sec.~\ref{sec:SecurityAnalysis}.

\subsubsection{TRANSMON Design: Data Protection Unit (DPU)}

The DPU forms another integral part of the TRANSMON, and its function is to provide data-level protection.
This is achieved by blocking: 1)~over-writing of sensitive data in the event of unauthorized writes to specific memory locations, or 2)~writing out particular data of sensitive nature.
The latter serves to protect \textit{soft assets}, e.g., private cryptographic keys, from leaking inadvertently into the system's shared memory, e.g.,
       by malicious ``shadow writes''~\cite{basak17}.

A write transaction is blocked when the DPU PRS contains a relevant policy that disables writing of particular, restricted data to
a specified address range.
Since DPU policy checks can only work during the data phase of the AHB-Lite protocol,
we have to
keep the data, address, and control signals all registered
until the check is completed;
this registering is done within the SAF.
Hence, the DPU incurs one additional cycle delay in all transactions related to write-restricted data, but all other transactions not covered by DPU policies are not delayed.

\begin{figure}[tb]
\centering
\includegraphics[width=\columnwidth]
{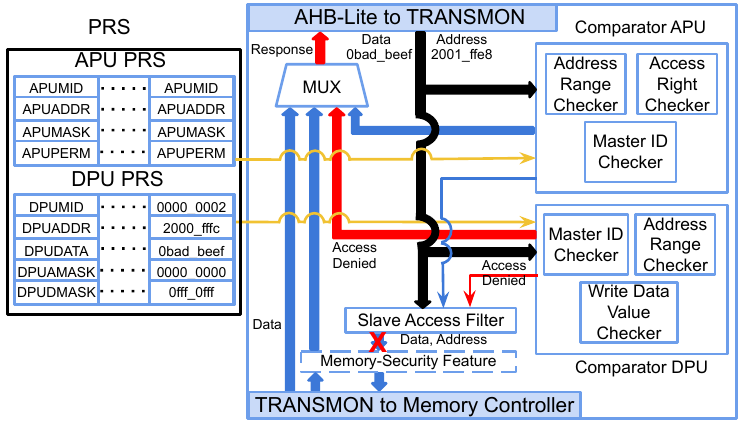}
\caption{A DPU policy in action in the TRANSMON.
Essentially, the core with ID 0x2 tries to write out a sensitive software asset, e.g., a private cryptographic key, with the value of 0xBAD\_BEEF, to a restricted memory region.
The access is blocked by the DPU from passing to the memory controller, and an error message is returned.
See also Sec.~\ref{sec:SecurityAnalysis} for more details.}
\label{fig:DPU_example}
\end{figure}

       As with the APU, recall that a DPU makes use of its own PRS.
Figure~\ref{fig:TRANSMON}
shows the five paramters of a DPU policy:
\begin{itemize}
\item DPUMID, which identifies the master whose write transaction is to be verified against this policy;
\item DPUADDR, a 32-bit memory address, which designates the address where the write permission is restricted;
\item DPUDATA, a 32-bit, write-restricted data value;
\item DPUDMASK, a 32-bit data mask; and
\item DPUAMASK, a 32-bit address mask.
\end{itemize}
An example for a DPU policy is illustrated in Fig.~\ref{fig:DPU_example}, and the related simulation is provided in Sec.~\ref{sec:SecurityAnalysis}.

\subsubsection{TRANSMON Design: Memory-Security Feature (Optional)}
\label{sec:mem_sec}

To protect against faults or malicious modifications within the shared system-level memories themselves,
schemes like ECC, cyclic redundancy check (CRC), data mirroring, or a combination of these can be implemented.
For example, an ECC implementation based on the well-known Hamming code would require four extra bits per memory byte, translating to 50\%
memory cost, and could
only serve to detect at most two corrupted bits per byte.
The advantage of ECC, however, is that it can be calculated during computation time without any latency overhead.
A CRC implementation is more suitable when the memory data to protect is not supposed to change, e.g., for a firmware/software
image. Also, CRC can be implemented with little additional circuitry.
Still, the CRC computation needs to be run for chunks of data intermittently; this may halt some regular computation and can thus impact the overall throughput.
In data mirroring, the data is simply copied into another (trusted) memory, which would naturally induce an overhead of 100\%.

We envision some memory-security scheme as
follows:
1)~the TRANSMON computes an ECC for any write-out that is allowed;
2)~the ECC result is stored in some separate and trusted control memory, whereas the actual data is stored in the shared-memory chiplet attached to the
TRANSMON; 3)~during
read-out, the TRANSMON validates the data using the stored ECC result.
If that check fails and cannot be corrected via the ECC, the data
is rejected and the related memory region is marked as \textit{tainted} and not used further.

Since address handling is covered by the AHB-Lite protocol,
one may implement this scheme such that ECC results are fetched in parallel, without inducing additional delays for read-out transactions.
We
note that ECC results have to be stored in a trusted, separate memory chiplet.
Finally, neither the above nor other memory-security features
can protect against \textit{erroneous data} arising from hardware/software failures or malicious activities. Such risks can only be mitigated at
the system level, e.g., by redundant computation and majority voting on results~\cite{
	mavroudis17,
	nguyen08}. Note that a multi-core architecture like ours can be readily
tailored for such needs, but related efforts are scope for future work.
        
\section{Physical Design}
\label{sec:PhysicalDesign}

\begin{figure}[tb]
\centering
\includegraphics[width=\columnwidth]{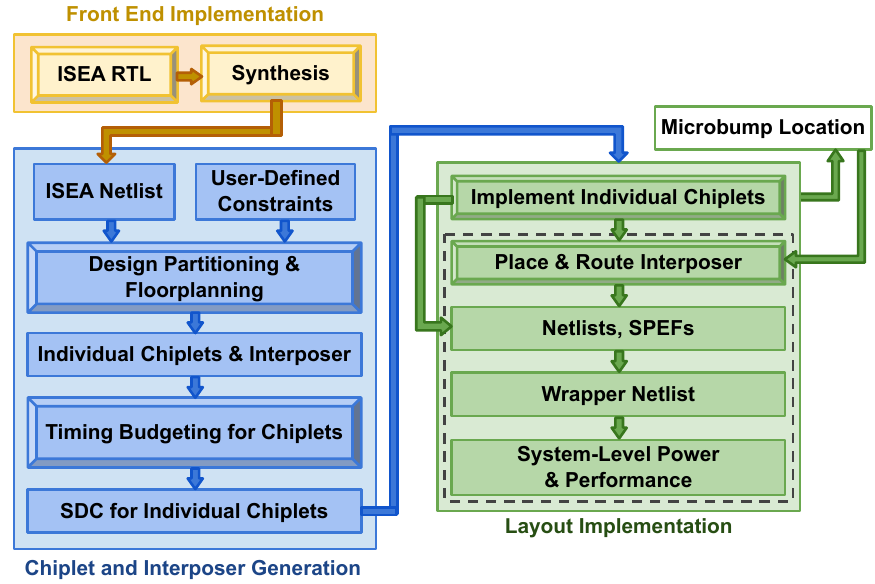}
\caption{Our end-to-end physical-design flow based on commercial tools.
After synthesis of ISEA,
the chiplets and interposer are floorplanned.
Next, all components are implemented, along with microbumps planning.
If chiplets are given as hard IPs, only the interposer
implementation is required (dashed box).
\label{fig:layout_flow}
}
\end{figure}

Next, we elaborate on our end-to-end physical-design flow, which is devised for our 2.5D RoT, but can also be applied for any other active 2.5D system.
The flow is illustrated in Fig.~\ref{fig:layout_flow}, and some highlights are discussed next.
Note that the flow is leveraging commercial tools, libraries, and technologies for all key design steps such as placement and routing or handling
	of timing constraints; see also Sec.~\ref{sec:exp_setup} for more details on the setup.
For the interested reader/designer, we would also provide access to our flow upon request.

First, for the front-end implementation, the whole ISEA register-transfer level (RTL) design is synthesized to obtain a \textit{full-system netlist}.
Chiplets provided as soft IP are to be synthesized here as well.

\subsection{Chiplets and Interposer Generation}

The full-system netlist is then partitioned into \textit{banks}, which simply represent the logic
and memory chiplets.
Our flow provides flexibility to the designer when choosing the number of logic/memory banks as needed for the chiplets organization.  Based on
the full-system netlist, we derive the timing budgets and obtain separate timing constraints (SDC files) for the individual chiplets and the
active interposer.

Next, we generate the full-system floorplan.
Relevant parameters are to be provided by the designer, such as utilization for individual chiplets and their aspect ratio, and they are used toward floorplanning of the
related core/memory banks.
Also, the floorplans of memory banks capture the placement of memory modules within each bank.
The designer is also required to provide
the arrangement of chiplets over the active interposer.
Finally, the interposer die outline is derived from the full-system floorplan.

All the floorplan data is kept in
\textit{Tool Command Language (TCL)} format,
which eases the use of a regular 2D implementation flow while designing the chiplets and the active interposer.
We emphasize that  our  flow is flexible with respect to accommodating chiplets that are either designed in-house or, what is more practical, procured  as  physical
hard IP  from commercial  vendors. 
For  such hard IP, the design steps are more straightforward and essentially
cover only the chiplets arrangement over the interposer and the design of the active interposer itself.  When  procuring such hard IP, it is easy to see that the designer
has no freedom for any intra-chiplet  optimization. Still, our flow allows the designer to explore different chiplet arrangements, which eases the system-level design space exploration along with an investigation of timing and power consumption.

\subsection{Layout Implementation}

For chiplets obtained as soft IP,
the related netlists 
have to proceed through a standard 2D implementation flow, to obtain the  individually placed-and-routed  chiplet layouts.
During this step, we also derive the locations for microbumps, which serve the physical connection between chiplets and the interposer.
Those microbumps are initially placed around the vicinity of drivers/sinks, while further on-track legalization is performed to avoid routability issues
and maximize the utilization of routing resources for the chiplets.
Thereafter, the microbump locations of all chiplets
	are used to define the microbump  locations  for  the  interposer.
Next, the RC parasitics for each chiplet are generated as SPEF files from their post-routed  layouts. Along
with the final
netlist, these SPEF files are used later on for sign-off analysis, i.e., to evaluate power consumption and timing.

Once the 2D implementation of all chiplets is completed---which is skipped in case chiplets are obtained as hard IP---the placement and
routing of the active interposer follow. First, the  interposer netlist is imported, which describes the AHB-Lite components, the HWSFs of ISEA,
	and
the pre-defined interposer microbump  locations.
Second, a 2D implementation of the active interposer follows.
We note that we do not engage in any cross-optimization between chiplets and interposer, which is essential for the scenario of chiplets obtained as hard IP.
Third, the RC parasitics for the active interposer design are extracted and exported along with
the final netlist, and the GDSII is streamed out. Finally, the RC parasitics for the microbumps are modeled into the SPEF file of a \textit{wrapper netlist}.

To evaluate the system-level power consumption and timing of ISEA, including all computing and memory chiplets, all
individual netlists and their SPEF files are used along the wrapper netlist with its own SPEF file.

\section{Experimental Evaluation}
\label{sec:experiments}
\subsection{Setup}
\label{sec:exp_setup}

The RTL code for the complete system, including the cores, AHB-Lite bus, TRANSMONs, \textit{etc.}, has been realized using \textit{Verilog}.
We release the license-free parts of the RTL~\cite{webinterface}. 
Synthesis was performed via \textit{Synopsys DC} and layout generation via \textit{Cadence Innovus v.17.10}.
Verification
and simulation runs
have been carried out via \textit{Synopsys VCS}.
The \textit{ARM IAR} suite has been used to compile \textit{C} code to run on ISEA.

We implement ISEA as 64-core \textit{ARM Cortex-M0} multi-chiplet system for a PoC.
As baseline, the 64 cores are organized into four computing chiplets, each holding 16 cores.  For another configuration, to study the impact of
	system organization on layout costs, we reorganize the 64 cores into eight chiplets, each holding eight cores.
Concerning security policies, our baseline configuration supports 16 APU and 16 DPU policies for each TRANSMON.  To study the impact of
policies being supported by TRANSMONs on layout costs, we also consider configurations with 32, 64, and 128 APU and DPU policies being supported
by each TRANSMON.

For both the computing and shared-memory chiplets, we leverage the commercial 65nm \textit{GlobalFoundries} technology 
and \textit{ARM} standard cell and memory libraries, representing the advanced but untrusted facility.
We employ four shared-memory chiplets with 1 MB SRAM each,
build up from 16 memories at 64 kB.
For the active interposer, we use the \textit{Synopsys SAED} 90nm technology,
representing the older but trusted facility.
For brevity, we also refer to both technologies as 65nm and 90nm, respectively.
Note that the 90nm technology
does not provide memory modules;
thus, we have to refrain from implementing any memory-security feature for this PoC, as we cannot provision for a separate, trusted memory chiplet
required for such features.
For both technologies, we use a supply voltage of 1.08 V, and we consider their respective slow corners. Note that doing so allows for heterogeneous 2.5D integration without the need for level shifters.
In reality,
the advanced but untrusted facility versus the older but trusted facility may support technology nodes that are further
apart, but
we were constrained in choices by the libraries available to us.
Microbumps connecting the interposer and chiplets have a width of 5$\mu m$ and a pitch of 10$\mu m$.
We utilize 7 metal layers for both the 90nm and the 65nm technology.

\subsection{Security Analysis}
\label{sec:SecurityAnalysis}

\begin{figure}[tb]
\centering
\includegraphics[width=.95\columnwidth]{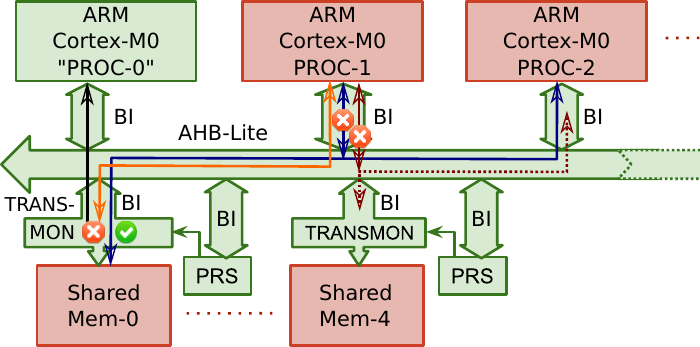}
\caption{\markup{Figure revised---}Various scenarios for system-level communication as handled within ISEA. Red crosses mean that the related threats are prevented,
	whereas a green check means that the transaction is approved.}
\label{fig:ISEA_examples}
\end{figure}

We study various scenarios for securing computation using ISEA.
First,
	we illustrate how critical threats
(i.e., snooping, spoofing, modifying, diverting, and
man-in-the-middle attacks) are ruled out by ISEA in the first place
(Fig.~\ref{fig:ISEA_examples}).
More specifically, there is an approved transaction between
PROC-2 and the shared memory with slave ID 0 (represented as blue arrow and green check in the TRANSMON of the memory).  At the same time,
PROC-1 seeks to snoop on that communication. This threat is blocked physically, directly by the BI (bus interface) of PROC-1, as the BI itself
delegates only data originating from/destined to PROC-1.  Next, PROC-1 tries to illicitly act as man-in-the-middle between PROC-2 and the
shared memory with slave ID 4 (represented as dashed, dark-red arrow). This threat is blocked directly at the BI as well---the BI
hard-codes the master ID 1 into any outgoing request, thereby preventing PROC-1 from masquerading its ID.
Finally, PROC-1 also tries to access some data in the shared memory with slave ID 0 (orange arrow).
However, this particular request is not approved by any policy and, thus, rejected. PROC-0 is informed about this blocked request
as well (black arrow).

\begin{figure*}[tb]
\hfill
\subfloat
{
	\includegraphics[width=.473\textwidth]{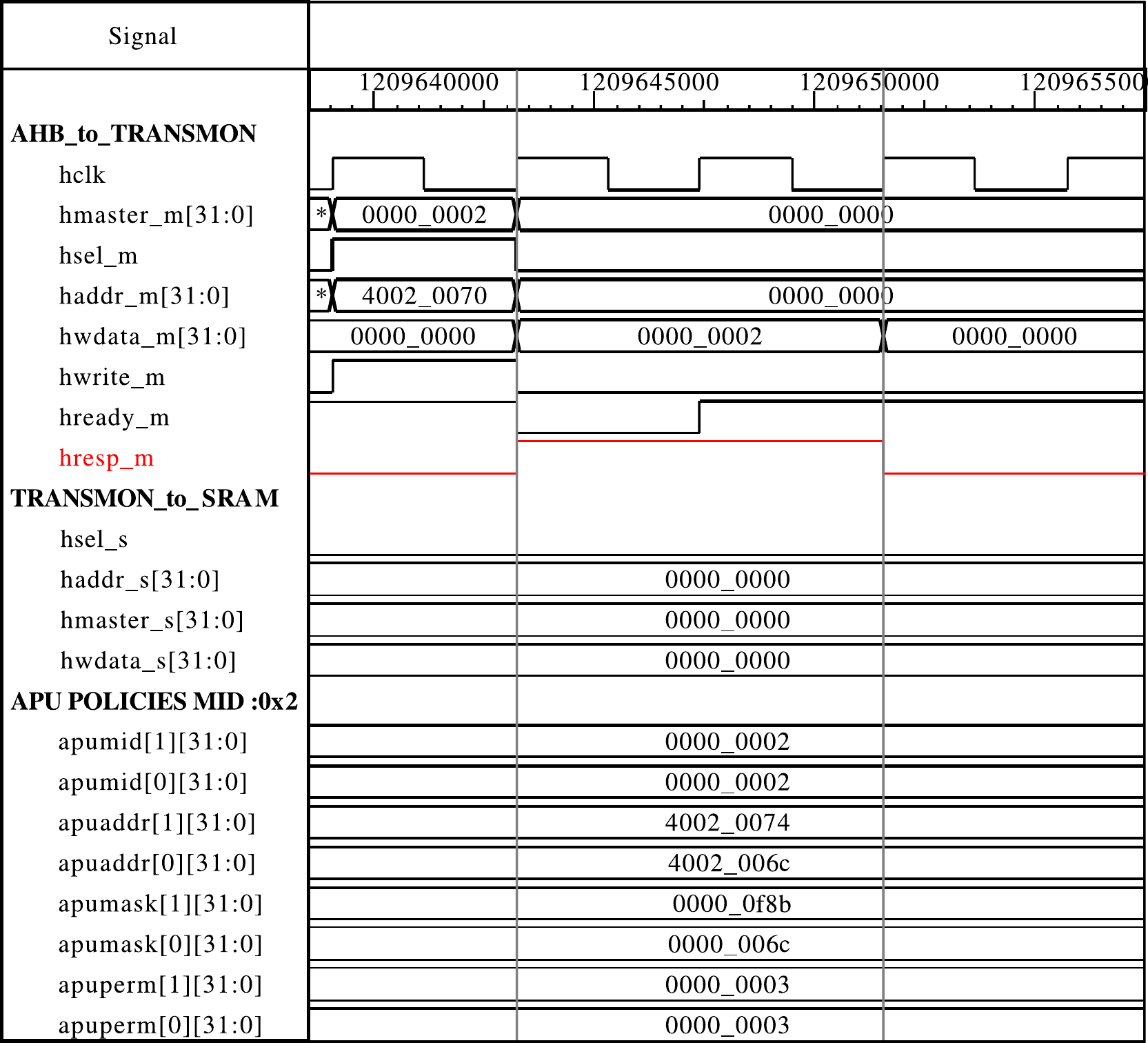}
}
\hfill
\subfloat
{
	\includegraphics[width=.506\textwidth]{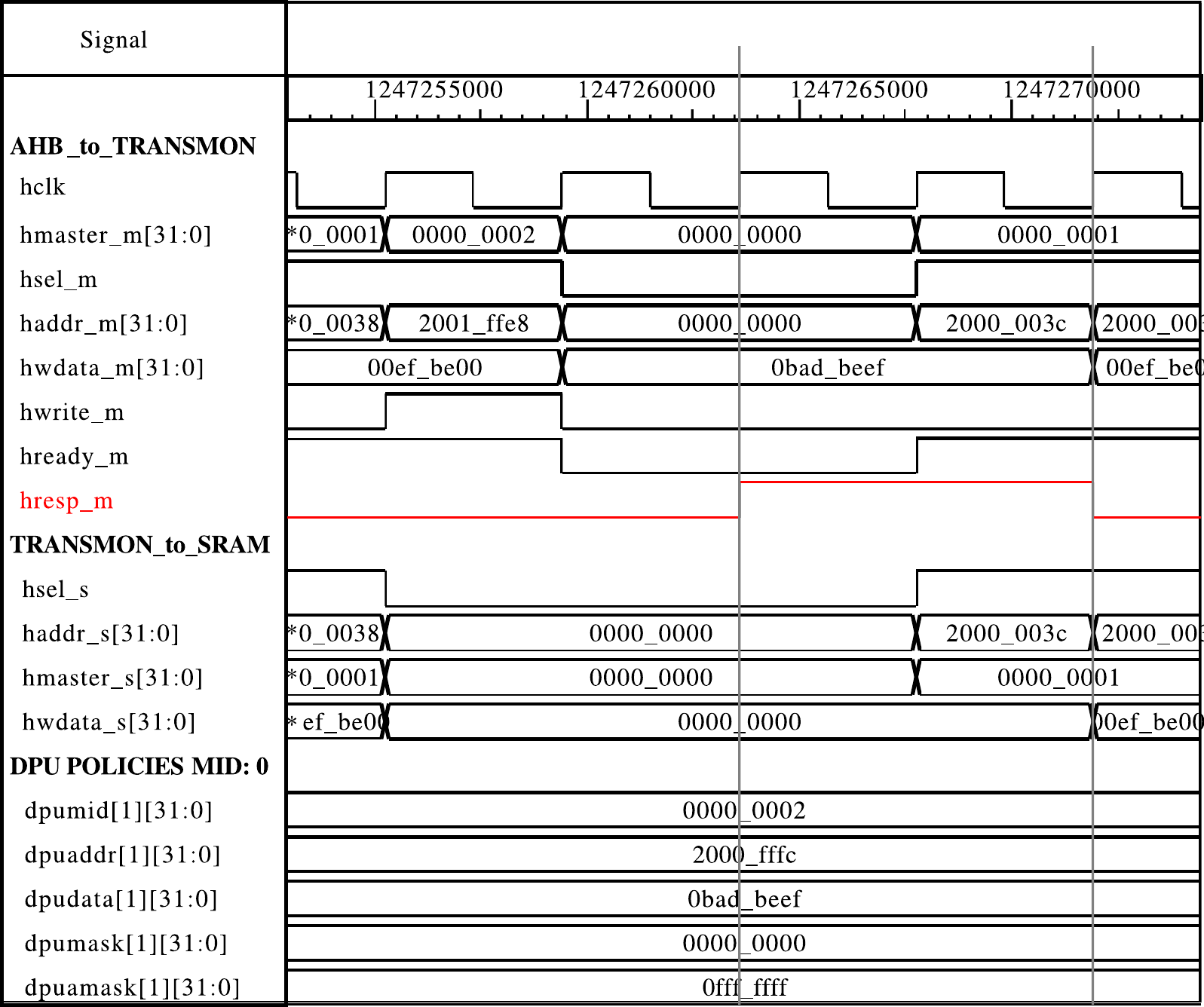}
}
\hfill
\caption{\markup{Figure revised---}Demonstration of ISEA at hardware simulation level, using \textit{Synopsys VCS}.
Left: malicious manipulation of data is blocked by an APU policy.
Right: unpermitted write-out of a secret key \textit{0XBAD\_BEEF} is blocked by a DPU policy.
Each blocking is receipted by the error message \textit{hresp\_m}.
AHB-Lite signals are described in Table~\ref{table:amba_ahb_signals}.
\label{fig:simulations}
}
\end{figure*}

Next, we explore various scenarios for runtime monitoring against malicious access or modification of the system-level shared memory.
These scenarios serve to show-case the working of ISEA in some detail, based on hardware simulation using \textit{Synopsys VCS} along with \textit{C} code compiled
for the Cortex-M0 cores using \textit{ARM IAR}.
\pending{AHB-Lite signals relevant for understanding of these simulations are listed in
Table~\ref{table:amba_ahb_signals}.}
Aside from the particular scenarios considered, the number of policies can increase and their interaction can become more complex in practice, depending
on the application(s) running on ISEA and the resulting security requirements.
We also note again that a full system-level software stack and related simulation efforts, also considering the orchestration and joint working
of PROC-0 and the TCU, is scope for future work.

\subsubsection{Protection of Memory Ranges}
Here, ISEA is tasked exemplarily with executing a fast Fourier transformation (FFT).
The FFT is an essential building block for many signal processing applications, and it can be parallelized 
straightforwardly.
As indicated in
Sec.~\ref{sec:ArcImplementation},
task scheduling is handled by the trusted control processor PROC-0, which
also arranges the input data within the system's shared memory space.
The FFT computations within each core is started upon receiving an interrupt from PROC-0, and
once the processing is done for all cores, the final results are gathered by PROC-0.
Toward that end, we implement custom interrupt handler for the cores, to perform computation as controlled by PROC-0.

The policies are compiled such that the intermediate FFT results calculated by one core cannot be modified by other, maliciously acting cores. That is, we protect the shared-memory regions assigned to each core via APU policies.
For example,
the core with ID 0x2 has access to the address range
0x4002\_0000
to 0x4002\_006C
and the range from 0x4002\_0074
to 0x4002\_0FFF,
but not to other addresses such as 0x4002\_0070 (i.e., where the
core with ID 0x1 stores its result).
Note that the address ranges are derived by the APU in an efficient manner, i.e., without need for complex comparator logic, using simple bit-wise operations.
For example for the related simulation in Fig.~\ref{fig:simulations}(left),
the start address
	0x4002\_0000 is \textit{APUADDR[1] AND NOT(APUMASK[1])} and
the end address
	0x4002\_006C is \textit{APUADDR[1] OR APUMASK[1]};
similarly, the start address
	0x4002\_0074 is \textit{APUADDR[2] AND NOT(APUMASK[2])} and
the end address
	0x4002\_0FFF is \textit{APUADDR[2] OR APUMASK[2]}.
	
As shown in the waveform in Fig.~\ref{fig:simulations}(left),
the core with ID 0x2 tries to access the
address 0x4002\_0070 to write out the data 0x0000\_0002.
Note that for AHB-Lite in general,
the address phase comes first and the data phase one cycle after.
The transaction is blocked by the APU, and the data in the memory remain protected and as is, indicated by the fact that the memory-controller signals are \textit{not} reflecting the requested write out.
At the same time, the error message \textit{hresp\_m} is returned.
Finally, note that this particular example is the same as in the conceptional Fig.~\ref{fig:APU_example}.

\begin{table}[t]
\begin{scriptsize}
\centering
\caption{Selected Signals for AHB-Lite}
\setlength{\tabcolsep}{1.0mm}
\begin{tabular}{c  p{7.4cm}}
\hline
{\bf Signal}&{\bf Description}\\
\hline
HCLK    & Bus clock; timing of all signals is related to the rising edge of HCLK.               \\ \hline
HMASTER & Master ID; a unique ID assigned to each master attached to the bus.  \\ \hline
HSEL    & Slave select; indicates that the current transaction is intended for the selected slave.\\ \hline
HADDR   & System address; identifies the address as related to the slave. \\ \hline
HWDATA  & Write data; used to transfer data from the master to the bus slaves during write operations and vice versa for read operations. \\ \hline
HWRITE  & Transfer direction; HIGH indicates a write transfer, whereas LOW indicates a read transfer.  \\ \hline
HREADY  & Transfer status; HIGH indicates that a transfer has finished on the bus; to extend the transaction, this signal is to be driven LOW. \\ \hline
HRESP   & Transfer response; provides feedback on the status of the transfer; used as receipt for security approval/rejection in our work. \\ \hline
\end{tabular}
\label{table:amba_ahb_signals}
\end{scriptsize}
\end{table}

\subsubsection{Protection of Private Assets}
Here, a malicious core tries to write out some soft asset, e.g., a private cryptographic key.
The DPU covers this kind of threat;
the related DPU policy concerns the actual data.

For Fig.~\ref{fig:simulations}(right), a DPU policy is set to track a write transaction by the
core with ID 0x2 to the restricted memory region between addresses 0x2000\_0000
to 0x2FFF\_FFFF,
concerning the sensitive data
0x0BAD\_BEEF.
Note that the DPU derives address ranges like the APU; the start address is
\textit{DPUADDR AND NOT(DPUAMASK)} and the end address is \textit{DPUADDR OR DPUAMASK}.
Also, the sensitive data is derived similarly, as \textit{DPUDATA AND NOT(DPUMASK)}.

The simulation waveform in Fig.~\ref{fig:simulations}(right) shows an attempt to write out the
restricted data value to address 0x2001\_FFE8, which is blocked.
Here as well, the error message \textit{hresp\_m} is returned.
Note that subsequently another, unrelated read transaction is approved, which can be seen
by the \textit{hready\_m} signal being turned on during the related data phase.
Finally, note that this particular example is the same as in the conceptional Fig.~\ref{fig:DPU_example}.

\subsubsection{Protection of Shared Assets}
Here, we assume that two or more cores require a \textit{semaphore} for software-based program and data control.
Semaphores can be stored in the SRS, the
shared register space, which is part of the 2.5D RoT, hence trustworthy by itself (Sec.~\ref{sec:ArcImplementation}).
Consider a maliciously acting core tries to over-write the semaphore to be able to access/execute data/program regions otherwise not accessible.
Here, a DPU policy is needed to monitor the actual data access to the semaphore, whereas a generic APU policy would not suffice.

\begin{figure}[tb]
\centering
\includegraphics[width=\columnwidth]{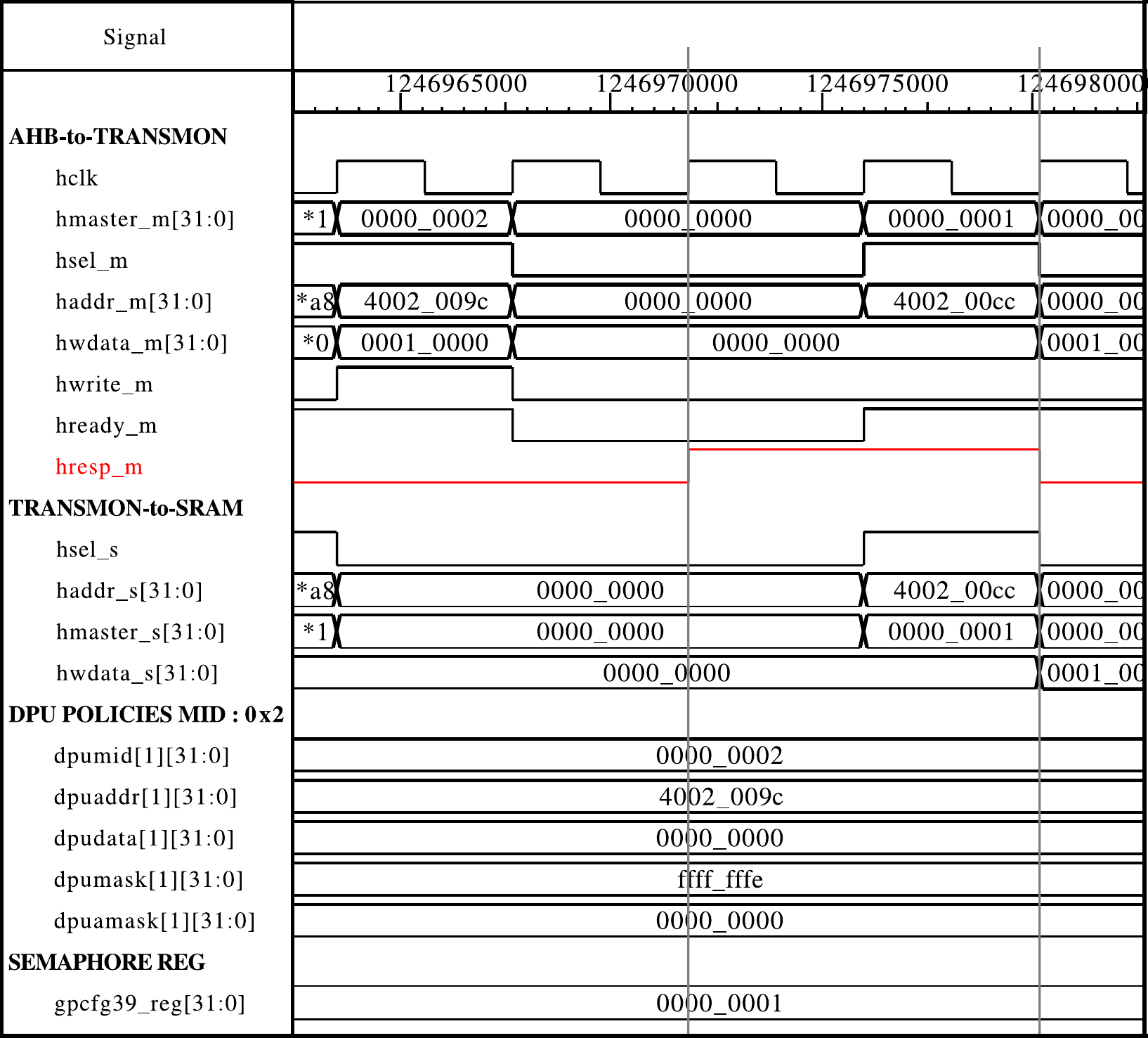}
\caption{\markup{Figure revised---}Demonstration of ISEA at hardware simulation level, using \textit{Synopsys VCS}.
Malicious over-writing of a semaphore in \textit{gpcfg39\_reg} is blocked by a DPU policy.
The blocking is receipted by the error message \textit{hresp\_m}.
AHB-Lite signals are described in Table~\ref{table:amba_ahb_signals}.
\label{fig:simulations_semaphore}
}
\end{figure}

Figure~\ref{fig:simulations_semaphore} shows how such a malicious transaction is blocked.
Here \textit{gpcfg39\_reg} is considered as a semaphore register.
For the core with ID 0x1, to obtain the ownership of this semaphore,
it has to write 0x0000\_0001 to the above register, but can do so only while the semaphore register value is 0x0000\_0000, i.e., while the semaphore is
available.
For the core with ID 0x2, it has to write 0x0000\_0010 to obtain the semaphore, and so on.
Naturally, one core should not be able to obtain the semaphore when it is already used by any other core---a DPU policy is compiled to implement this restriction.
In the simulation, the policy is set for the core with ID 0x02 to prevent any malicious writing of ``0'' to the last bit (\textit{DPUDATA AND
		NOT(DPUMASK)}) of the semaphore register. The waveform shows such an attempt to clear that last bit, which is
blocked, along with the error message \textit{hresp\_m} being returned.
Note that the waveform shows subsequently another, unrelated read transaction initiated by the core with ID 0x01, which is approved.

\subsection{Layout Analysis}
\label{sec:layout_analysis}

\begin{figure*}[tb]
\hfill
\subfloat
{
\includegraphics[height=63mm]{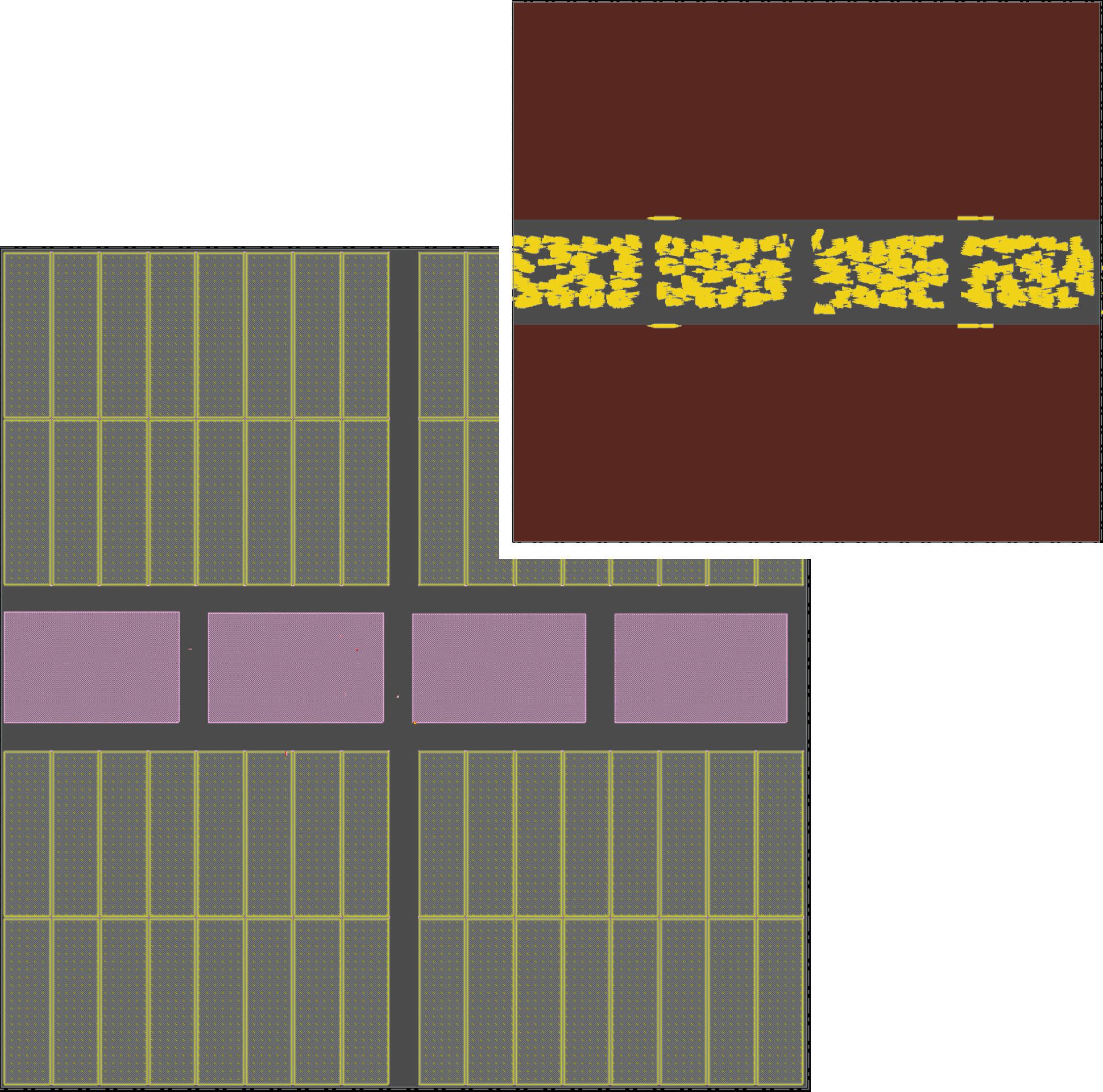}
}
\hfill
\subfloat
{
\includegraphics[height=63mm]{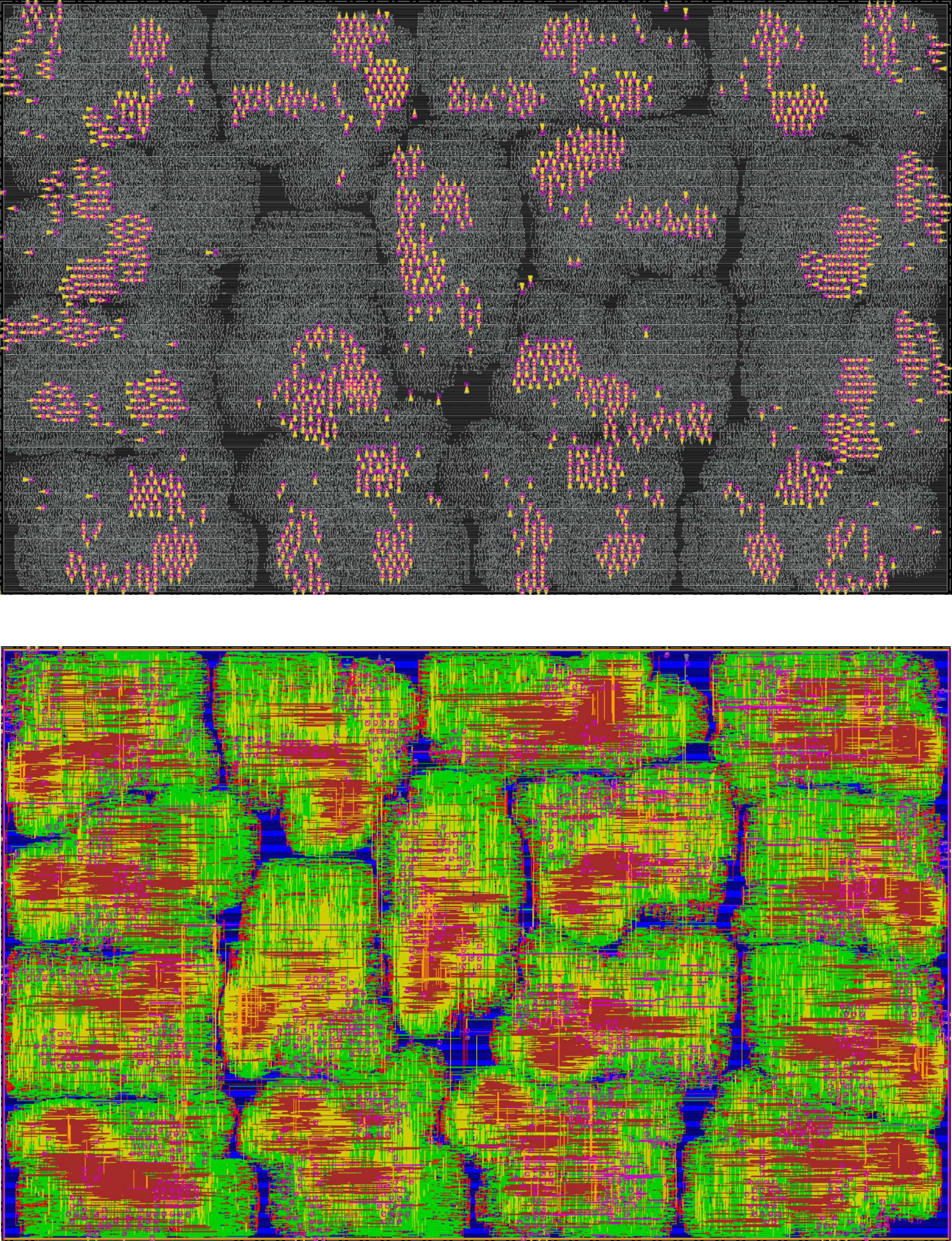}
}
\hfill
\subfloat
{
\includegraphics[height=63mm]{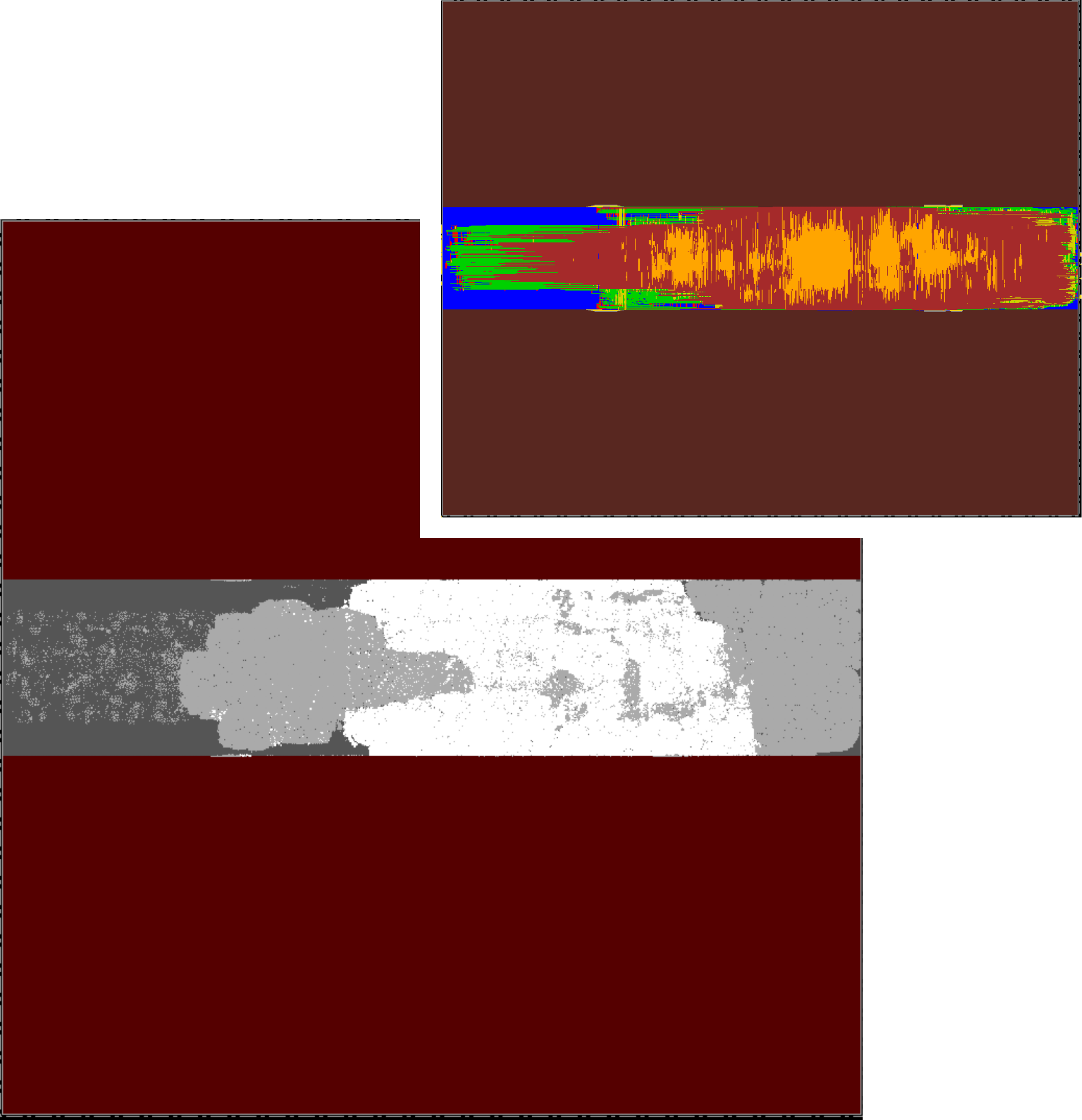}
}
\caption{Layout snapshots for the 64 multi-core, multi-chiplet baseline version of ISEA, obtained using our 2.5D design flow.
		For visual clarity, the power-distribution networks are not shown for the routing snapshots.
Left: floorplan of ISEA. Four ARM chiplets in the middle and four shared-memory chiplets around;
inset: interposer microbump locations.
Center: One ARM Cortex-M0 chiplet, with 16 cores. Logic and microbump locations (top), routing (bottom).
Right: ISEA implementation in the active interposer.
TRANSMON and other HWSFs in white, regular AHB-Lite components in grey. Placement regions are constrained, for cost-efficient interposers,
	 e.g., using \textit{Intel's EMIB} technology;
inset: interposer routing.
\label{fig:layouts}
}
\end{figure*}

Using our 2.5D design flow, we investigate the physical layouts of various ISEA configurations. In Fig.~\ref{fig:layouts}, we provide snapshots for the baseline 64-core 2.5D version of ISEA.
We note that the details discussed below are based on our commercial-grade implementation setup (Sec.~\ref{sec:exp_setup}).

\subsubsection{2D Implementation, ISEA in General}

First, we analyze the impact of our security-enforcing features on layout costs.
To do so, we compare the baseline ISEA multi-core design to a corresponding but \textit{non-secure} design, both
implemented via a regular 2D IC flow using the \textit{GlobalFoundries} 65nm technology.
For the non-secure design, we maintain all Cortex-M0 cores, memories, AHB-Lite components, and glue logic, but we drop all HWSFs
such as TRANSMONs, PRSs, \textit{etc.}
\pending{We note that, from a conceptional point of view, using the 
\textit{Synopsys SAED} 90nm technology would be more apt, as this technology was designated as the trusted node.
Then, the corresponding secure 2D implementation would represent
the system as implemented exclusively using the trusted technology.
However, given that the
90nm technology does not provide memory modules, we have to resort to the
65nm technology. For this reason, we also refrain from directly comparing the secure 2D implementation with the
secure 2.5D system later on.}

Table~\ref{tab:secure_non_secure_2D_65nm_PPA} provides the results for the 2D implementation.
For the secure design, we observe a
5\% reduction in critical delay and a 13.86\% increase in power consumption.
Note that we achieve a competitive critical delay for the secure design by breaking longer paths using \textit{pipelining}.
An increase in standard-cell area (2.48\%), instance count (29.57\%), buffer count (18.46\%), wirelength (31.49\%), and total capacitance
(35.44\%) are all expected, due to the proposed HWSFs (including all registers required for storage of policies, \textit{etc.})
and due to pipelining.
The die outline remains as is, however; no additional silicon cost occurs.
These results provide the range of costs to be expected for ISEA, that is at least for this particular PoC implementation.

\begin{table}[tb]
\centering
\caption{2D Implementation Results for Non-Secure Versus Secure Designs, Both in \textit{GlobalFoundries} 65nm}
\label{tab:secure_non_secure_2D_65nm_PPA}
\begin{tabular}{ccc}
\hline
\textbf{Metrics} 
& \textbf{Non-Secure (2D)} 
& \textbf{Secure (2D)} 
\\ \hline
Critical Delay ($ns$) 
& 9.79 & 9.29
\\
Power Consumption ($mW$) 
& 239.5 & 272.7 
\\
Standard-Cell Area ($\mu m^2$)
& 24,127,403 & 24,725,036 
\\
Total Die Area ($\mu m^2$)
& 31,996,800 & 31,996,800 
\\
Total Instance Count
& 600,729 & 778,393 
\\
Total Buffer Count
& 132,477 & 156,929
\\
Total Wirelength ($m$)
& 28.9 & 38.2
\\
Total Capacitance ($nF$)
& 7.9 & 10.7
\\ \hline
\end{tabular}
\end{table}

\subsubsection{2.5D Implementation}

Table~\ref{tab:secure_2.5D_90nm_65nm_PPA} provides the physical-design results for the 2.5D baseline implementations.
As indicated, computing and memory chiplets are implemented using the 65nm technology
and the active interposer using the
90nm technology, respectively.
Here we also compare a secure design with a non-secure design; both contain the same set of computing and memory chiplets, and both hold
all AHB-Lite components in the active interposer, whereas the secure design further holds the proposed HWSFs in the interposer.

\begin{table}[t]
\centering
\caption{2.5D Implementation Results for Non-Secure Versus Secure Designs, Chiplets in \textit{GlobalFoundries} 65nm, Interposer in \textit{Synopsys SAED} 90nm}
\label{tab:secure_2.5D_90nm_65nm_PPA}
\begin{tabular}{ccc}
\hline
\textbf{Metrics} 
& \textbf{Non-Secure (2.5D)} 
& \textbf{Secure (2.5D)} 
\\ \hline 
Critical Delay ($ns$) 
& 9.72 & 9.83
\\
Power Consumption ($mW$) 
& 266.4 & 300.9 
\\
Standard-Cell Area ($\mu m^2$)
& 24,588,292 & 26,844,473
\\
Total Die Area ($\mu m^2$)
& 33,641,866 & 33,641,866
\\
Interposer Die Area ($\mu m^2$)
& 6,237,600 &  6,237,600
\\
Total Instance Count
& 569,574 & 745,693
\\
Interposer Instance Count
& 69,742 & 249,085
\\
Total Buffer Count
& 141,151 & 169,344
\\
Total Wirelength ($m$)
& 30.5
& 40.5 \\
Total Capacitance ($nF$)
& 7.92 & 10.89
\\ \hline
\end{tabular}
\end{table}

For the secure design, we observe an overhead of 1.13\% for critical delay, 12.95\% for power consumption, and 32.79\% for wirelength, respectively.
The standard-cell area is increased by 9.18\%, while instance count and buffer counts are increased by 30.92\% and 19.97\%, respectively.
As before, these costs are attributed to the HWSFs (including all PRS registers, \textit{etc.}), but here the costs are further impacted by the
migration to 2.5D and by the heterogeneous technology setup. More specifically, due to the migration to 2.5D, all the system-level interconnects are
now passing through the active interposer, with all chiplets connected to this fabric through microbumps.
Thus, timing closure for the interposer is subject to the multiple chiplets, which requires more effort.
More importantly even, recall that the active interposer is implemented in the older 90nm technology. Therefore, higher costs are naturally to be
expected, especially for all the HWSFs residing in the interposer.
As with the 2D designs, there is no impact on the die areas for the 2.5D designs. In fact, the size of the interposer is dominated by the size and
arrangement of the chiplets mounted on top of it, not by the standard-cell area of the additional logic incurred for the HWSFs within the interposer.

We emphasize again that we refrain from any cross-optimization between chiplets and the active interposer, to account for the practical assumption of hard-IP chiplets obtained as commodity components from the open market.
Moreover, we note that our flow allows the designer to constrain the active area of the interposer (Fig.~\ref{fig:layouts}(right)) and
the placement of microbumps (Fig.~\ref{fig:layouts}(left)).  Doing so enables
the final vendor to
manufacture only a small CMOS chip for the interposer, instead of the whole outline, which naturally helps save commercial cost.
Such a small chip could be supported by
Intel's EMIB technology~\cite{
		EMIB19}.

We note that the results above are all subject to the ISEA PoC baseline configuration, i.e., 64 cores are organized into four computing
chiplets, each holding 16 cores, and 16 APU and 16 DPU policies are supported by each TRANSMON. To understand the scaling of layout costs
incurred by the proposed HWSFs, we next conduct the following experiments:
\begin{enumerate}
\item We explore the impact
for the number of policies being supported, by re-implementing the active interposer for 32, 64, and 128
APU and DPU policies being supported by each TRANSMON;
\item We explore the impact of the system-level organization, by
rearranging the 64 cores into eight computing chiplets with eight cores each and re-implementing the whole system.
\end{enumerate}

\begin{figure}[tb]
\centering
\includegraphics[width=\columnwidth]{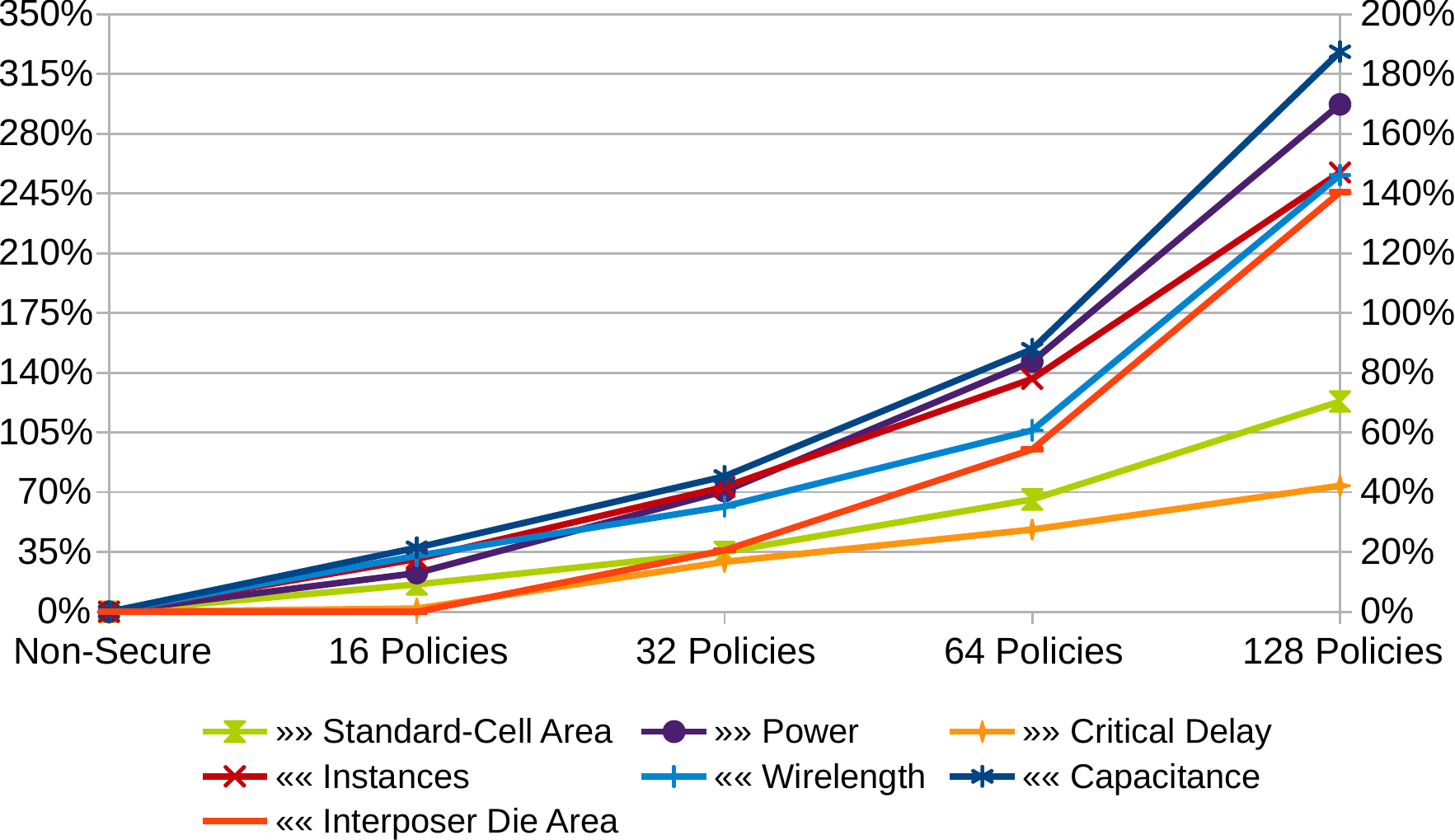}
\caption{\markup{Figure revised---}Scaling of layout costs for the 2.5D RoT, for varying numbers of APU and DPU policies being supported by each TRANSMON.
	The legend indicates by ``««'' or ``»»'' the assignment of each metric to the left or the right y-axis, respectively.
	Each metric is normalized to its respective value obtained for the non-secure 2.5D design.
\label{fig:layout_costs_policies}
}
\end{figure}

The results for 1) are illustrated in Fig.~\ref{fig:layout_costs_policies}. We note that layout costs are scaling up as expected. More
	specifically, for each doubling of the number of policies being supported, most metrics are approximately doubled in cost as
		well (considering their respective baseline cost for the initial configuration supporting 16 policies).
However, critical delays increase only linear---this
indicates that the physical designs are well optimized in terms of performance/timing paths.

The results for 2) are as follows: the standard-cell area is increased by 0.99\%, interposer die area by 47.35\%, power by 2.13\%, critical
	delay by -10.15\% (i.e., reduced by 10.15\%), total instance count is increased by 6.24\%, total wirelength by 9.41\%, and total
		capacitance by 6.24\%, respectively.  Given the reorganization of cores into double the number of computing chiplets, such costs
		are expected. More specifically, on the one hand, having to accomodate double the computing chiplets imposes a larger outline for the
		interposer. This is because the chiplets are not halved in size, as microbumps are dominating their outlines, and not the
		logic within.
		Due
		to the larger interposer die outline, we also observe larger total wirelength, along with more instances (required for buffering),
	higher capacitance, and marginally higher power consumption. On the other hand, the critical delay can be improved, thanks to some
		critical paths becoming shorter within the smaller computing chiplets, as well as due to the rearrangment of chiplets on top of the
		active interposer.

In short, these two experiments show reasonable costs, manifesting the practicality of our architecture for different configurations.
We also like to argue that, for other systems beyond our Cortex-M0 PoC implementation, e.g., when using \textit{RISC-V} cores
instead, the layout costs
might be better amortized over the respectively larger system.

Finally, we reiterate the fact that prior art is conceptionally different from ours.
Prior art embedded their HWSFs 
monolithically within 2D ICs, whereas the risks related to adversaries in the IC supply chain have
been largely overlooked.
For the few studies considering HWSFs along with 2.5D/3D integration for protection at runtime
(Table~\ref{tab:prior_art}),
recall that their security promises are most often still subject to the design and manufacturing of the whole chip and the trust---or rather lack thereof---into the related facilities, e.g., as it is the self-declared case with~\cite{valamehr13}. Therefore, we argue that a comparative study of ours
with prior art is neither meaningful nor practical and, hence, not provided here.

\section{Conclusions and Future Work}
\label{sec:conclusion}

We demonstrated a hardware security concept that provides a stringent physical separation, directly at the system level, between untrusted commodity
components and trusted security-enforcing components.
Our concept is in notable contrast to prior art where HWSFs are embedded monolithically in the same 2D IC as all other 
untrusted components and, thus, become inevitably subject to the trustworthiness of the design and manufacturing stages of that 2D IC.

For the first time, our architecture, dubbed ISEA, 
uses an active interposer as physically separate 2.5D root of trust, encompassing all proposed HWSFs and the system-level interconnect
fabric. 
ISEA is based on stringent policy-based verification of every bus transaction, and serves to protect the system from various software- or
hardware-emanating attacks.
We provide the license-free parts of our proof-of-concept implementation, which is
based on the \textit{Cortex-M0} core and the
\textit{AHB-Lite} bus system by \textit{ARM},
in~\cite{webinterface}.

Our work establishes trustworthy computation in the face of untrusted commodity components integrated into a
larger system, while maintaining the good economics of outsourced supply chains.
In fact, the security-enforcing vendor only has to focus on the HWSFs and can integrate commodity chiplets as needed, while
the system-level security remains intact even in the presence of any malicious behavior introduced by such
chiplets.
We believe that ISEA
   empowers secure computation by
construction, while maintaining scalability and flexibility for various systems. 

ISEA was tested at hardware simulation level under various threat scenarios and conceptionally demonstrated to offer robust security from malicious activities. Next, using
our proposed, commercial-grade 2.5D physical-design flow, we explored the practical scenario of integrating hard-IP chiplets on an active
interposer, with different technology nodes used for chiplets and the interposer.

For future work, we plan to extend ISEA for other system implementations, in particular with \textit{RISC-V} cores. We also plan for
system-level software implementation and simulation of ISEA, e.g., using \textit{gem5}.
We also plan to leverage our scheme for multi-party computation which, by definition, requires a root of trust for system-spanning security.
Moreover, we plan to apply and study the notions of redundant computation and majority voting for critical applications.

In the longer term, we envision a holistic 2.5D root of trust,
where we see the active interposer being augmented with side-channel sensors, e.g., to track power consumption of individual chiplets. As such, we would seek to track malicious activities which are more stealthy and refrain from targeting directly at system-level memory data.

\section*{Acknowledgments}
\label{sec:acknowledgments}

This work was supported in part by the Center for Cyber Security at NYU New York/Abu Dhabi (NYU/NYUAD) and by the NYUAD REF scheme under grant RE218.
The work of S.\ Patnaik was supported by the Global Ph.D.\ Fellowship at NYU/NYUAD.


\begin{IEEEbiography}[{\includegraphics[width=1in,height=1.25in,clip,keepaspectratio]{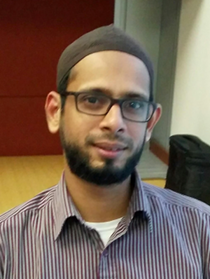}}]{Mohammed Nabeel} is a Chip Design Engineer from India. 
He received his Bachelors degree in electrical and electronics engineering from National Institute of
Technology--Calicut, India.

Mr.\ Nabeel is currently working as a Research Engineer at Center for Cyber Security at New York University Abu Dhabi (CCS-NYUAD). Apart from
working on research in the field of hardware security, he also focuses on implementing and prototyping the research ideas in Chip. He has around
12 years of industry experience in chip design -- specialized in Micro architecture, protocol know-how, RTL design, Synthesis, Static Timing
Analysis and post silicon bring up. Prior to joining CCS-NYUAD, he worked at Texas Instruments, where he worked on chips targeted for IoT and
Automotive and prior to that was with Qualcomm, where he worked on chips targeted for mobile phones and data cards.  He has around 10 conference
and journal papers and 1 issued US patent.
\end{IEEEbiography}

\begin{IEEEbiography}[{\includegraphics[width=1in,height=1.25in,clip,keepaspectratio]{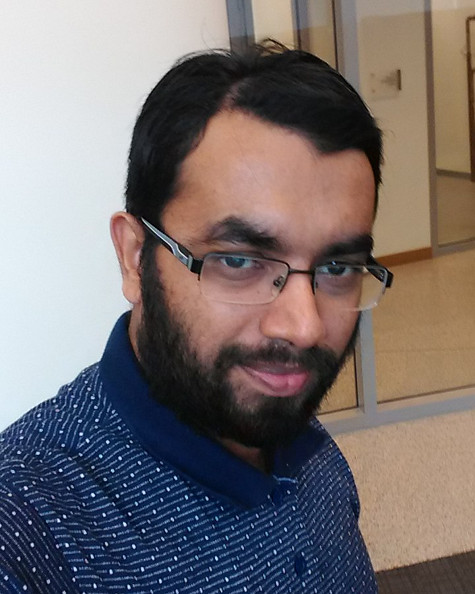}}]{Mohammed Ashraf}
received the bachelor’s degree
in electronics and telecommunication engineering from the College of Engineering Trivandrum,
Thiruvananthapuram, India, in 2005.

He is a Senior Physical Design Engineer from
India. 
He carries an experience of ten years in
the VLSI industry. 
He has worked with various
multinational companies like NVIDIA Graphics,
Santa Clara, CA, USA, Advanced Micro Devices,
Santa Clara, and Wipro Technologies, Bengaluru,
India. 
He worked also with Dubai Circuit Design,
Dubai Silicon Oasis, Dubai, United Arab Emirates. 
He is currently a Research Engineer with
the Center for Cyber Security, New York University Abu Dhabi, United Arab Emirates.
His work focus on the Physical Design/Implementation of the ARM Cortex M0 processor and its four secure variants.
\end{IEEEbiography}

\begin{IEEEbiography}[{\includegraphics[width=1in,height=1.25in,clip,keepaspectratio]
{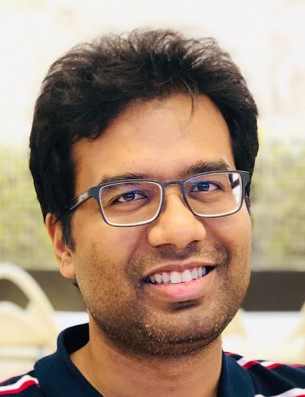}}]{Satwik Patnaik} (Graduate Student Member, IEEE) received
the B.E.\ degree in electronics and telecommunications from the University of Pune, India, and
the M.Tech.\ degree in computer science and engineering with a specialization in VLSI design from the Indian Institute of Information Technology and Management, Gwalior, India. 
He is currently pursuing the Ph.D.\ degree
with the Department of Electrical and Computer
Engineering, Tandon School of Engineering, New
York University, Brooklyn, NY, USA. 

He is also a Global Ph.D.\ Fellow with New York
University Abu Dhabi, United Arab Emirates. 
His current research interests include
hardware security, trust and reliability issues for CMOS and emerging devices
with particular focus on low-power VLSI Design.

Mr. Patnaik received the Bronze Medal in the Graduate Category at the
ACM/SIGDA Student Research Competition held at ICCAD 2018, and the Best Paper Award at the Applied Research Competition held in Conjunction
With Cyber Security Awareness Week, in 2017.
\end{IEEEbiography}

\begin{IEEEbiography}[{\includegraphics[width=1in,height=1.25in,clip,keepaspectratio]{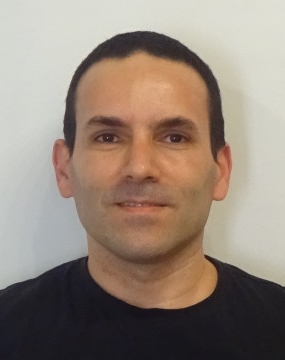}}]{Vassos Soteriou}
(Senior Member, IEEE) received the B.S.\ and Ph.D.\ degrees in electrical engineering from Rice University, Houston, TX, in 2001, and
Princeton University, Princeton, NJ, in 2006, respectively. 
He is currently an Associate Professor at the Department of Electrical Engineering,
Computer Engineering and Informatics at the Cyprus University of Technology. 
He is a recipient of a Best Paper Award at the 2004 IEEE International Conference on Computer Design. 
His research interests lie in high-performance computing, multicore computer architectures, and on-chip networks.
\end{IEEEbiography}

\begin{IEEEbiography}[{\includegraphics[width=1in,height=1.25in,clip,keepaspectratio]{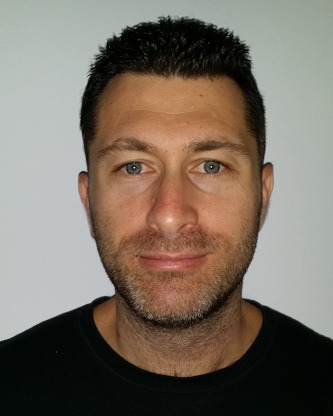}}]{Ozgur Sinanoglu} (Senior Member, IEEE)
received the first B.S.\ degree in electrical and electronics engineering and the second B.S.\ degree in computer engineering from Boğaziçi University, Istanbul, Turkey, in 1999, and the M.S.\ and Ph.D.\ degrees in computer science and engineering from the University of California at San Diego, CA, USA, in 2001 and 2004, respectively.  

He is a Professor of electrical and computer engineering with New York University Abu Dhabi (NYU Abu Dhabi), United Arab Emirates. 
He has industry experience with TI, Dallas, TX, USA, IBM, Armonk, NY, USA, and Qualcomm, San Diego, CA, USA. 
He has been with NYU Abu Dhabi since 2010, where he is the Director of the Design-for-Excellence Lab. 
His recent research in hardware security and trust is being funded by U.S. National Science Foundation, U.S. Department of Defense, Semiconductor Research Corporation, Intel Corp, and Mubadala Technology. 
His research interests include design-for-test, design-for-security, and design-for-trust for VLSI circuits, where he has more than 180 conference and journal papers, and 20 issued and pending U.S. Patents. 
He has given more than a dozen tutorials on hardware security and trust in leading CAD and test conferences, such as DAC, DATE, ITC, VTS, ETS, ICCD, and ISQED.

Prof.\ Sinanoglu won the IBM Ph.D.\ Fellowship Award Twice during his Ph.D.\ 
He is also the recipient of the Best Paper Awards at IEEE VLSI Test Symposium 2011 and the ACM Conference on Computer and Communication Security 2013. 
He was a (Guest) Associate Editor for the \textsc{IEEE Transactions on Information Forensics and Security}, 
the \textsc{IEEE Transactions on Computer-Aided Design of Integrated Circuits and Systems}, the \textit{ACM Journal on Emerging Technologies in Computing Systems}, the 
\textsc{IEEE Transactions on Emerging Topics in Computing}, \textit{Microelectronics Journal} (Elsevier), the \textit{Journal of Electronic Testing: Theory and Applications}, and \textit{IET Computers and Digital Techniques} journals. 
He is serving as the Track/Topic Chair or Technical Program Committee Member in about 15 conferences.
\end{IEEEbiography}

\begin{IEEEbiography}[{\includegraphics[width=1in,height=1.25in,clip,keepaspectratio]{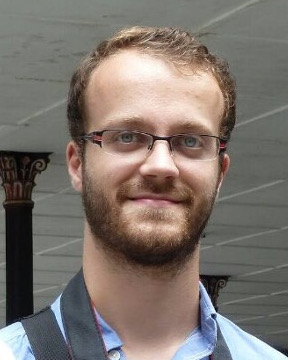}}]{Johann Knechtel}
(Member, IEEE) received the M.Sc.\ degree in Information Systems Engineering (Dipl.-Ing.) in 2010 and the Ph.D.\ degree in Computer Engineering
(Dr.-Ing., summa cum laude) in 2014, both from TU Dresden, Germany.
He is a Research Scientist with New York University Abu Dhabi, United Arab Emirates.
From 2015 to 2016, he was a Postdoctoral Researcher with the Masdar Institute of Science and Technology, Abu Dhabi;
from 2010 to 2014, he was a Ph.D.\ Scholar and Member with the DFG Graduate School on ``Nano- and Biotechnologies for Packaging of Electronic
Systems'' hosted at TU Dresden;
in 2012, he was a Research Assistant with the
Chinese University of Hong Kong, Hong Kong;
and in 2010, he was a Visiting Research Student with the
University of Michigan at Ann Arbor, MI, USA.
His research interests cover VLSI physical design automation, with particular focus on emerging technologies and hardware security.
He has (co-)authored around 50 publications.
\end{IEEEbiography}

\end{document}